%% file: main.tex
\documentclass{article}

\usepackage{arxiv}

\usepackage[utf8]{inputenc}
\usepackage[T1]{fontenc}
\usepackage[dvipsnames]{xcolor}
\definecolor{mycolor}{RGB}{14,3,123}
\usepackage[
    colorlinks=true,
    citecolor=blue,
    linkcolor=red,
    urlcolor=magenta,
    filecolor=cyan,
    menucolor=red,
    runcolor=filecolor,
    allcolors=mycolor
]{hyperref}
\usepackage{url}
\usepackage{booktabs}
\usepackage{amsfonts}
\usepackage{nicefrac}
\usepackage{microtype}
\usepackage{amsmath}
\usepackage{amssymb}
\usepackage{bm}
\usepackage{epstopdf}
\usepackage{algorithm}
\usepackage{algpseudocode}
\usepackage{float}
\usepackage{subcaption}
\usepackage{multirow}
\usepackage{tikz}
\usetikzlibrary{positioning,arrows.meta,bending,calc}
\DeclareGraphicsExtensions{.pdf,.png,.jpg,.eps}
\DeclareMathOperator{\diag}{diag}
\usepackage{cleveref}
\usepackage{lipsum}
\usepackage{graphicx}
\usepackage{natbib}
\usepackage{doi}

\input{macros}

\usepackage{etoolbox}
\let\oldresizebox\resizebox
\newcommand{\shrinkresizebox}[3]{\ifnum\value{table}=7 \oldresizebox{331.588pt}{#2}{#3}\else\oldresizebox{\ifdim\width>#1 #1\else\width\fi}{#2}{#3}\fi}
\AtBeginEnvironment{table}{\let\resizebox\shrinkresizebox}
\floatstyle{plaintop}
\restylefloat{table}

\title{Active Pilot Sampling for Multi-fidelity Estimation in Stochastic Optimization}

\date{October 04, 2026}

\newif\ifuniqueAffiliation
\uniqueAffiliationtrue

\ifuniqueAffiliation
\author{ \href{https://orcid.org/0000-0001-6764-9275}{\includegraphics[scale=0.06]{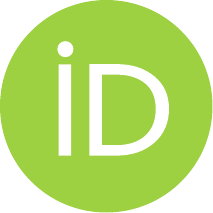}\hspace{1mm}Thomas E. Coons}\thanks{Corresponding author.}\\
	Department of Mechanical Engineering\\
	University of Michigan\\
	Ann Arbor, MI 48109 \\
	\texttt{tcoons@umich.edu} \\
	\And
	\href{https://orcid.org/0000-0003-3379-7514}{\includegraphics[scale=0.06]{orcid.pdf}\hspace{1mm}Aniket Jivani}\thanks{This work was conducted while at the Department of Mechanical Engineering, University of Michigan, Ann Arbor, MI 48109.} \\
        Viterbi School of Engineering\\
	University of Southern California\\
	Los Angeles, CA 90089 \\
    \texttt{ajivani@usc.edu}
    \And
	\href{https://orcid.org/0000-0001-6544-2764}{\includegraphics[scale=0.06]{orcid.pdf}\hspace{1mm}Xun Huan}\footnotemark[1]\thanks{\url{https://uq.engin.umich.edu/}} \\
    Department of Mechanical Engineering\\
	University of Michigan\\
	Ann Arbor, MI 48109 \\
    \texttt{xhuan@umich.edu}
}
\fi

\renewcommand{\shorttitle}{Active Pilot Sampling for Multi-fidelity Stochastic Optimization}
\hypersetup{
pdftitle={Active Pilot Sampling for Multi-fidelity Estimation in Stochastic Optimization},
pdfauthor={Thomas E.~Coons, Aniket~Jivani, Xun~Huan},
pdfkeywords={Monte Carlo, approximate control variates, optimization under uncertainty, covariance emulation, active learning, Bayesian optimal experimental design},
}

\begin{document}
\maketitle

\begin{abstract}
Stochastic optimization with high-fidelity nonlinear models is often computationally prohibitive because each objective evaluation requires sampling-based uncertainty quantification inside an outer optimization loop. Multi-fidelity estimators, such as approximate control variates (ACV), can reduce this cost by exploiting cheaper low-fidelity models, but they require model-output covariance information to determine control-variate weights and sample allocations. These covariances are generally unknown and must be estimated through pilot sampling. Existing pilot-sampling strategies primarily address one-off estimation and do not exploit the structure of iterative stochastic optimization, where covariance information may vary smoothly across the design domain and can be shared across optimization queries. We propose an active pilot-sampling framework that learns a probabilistic emulator of the model-output covariance matrix using a positive-definite parameterization based on matrix-logarithm-transformed correlations and log standard deviations. The emulator provides design-specific covariance estimates for constructing ACV estimators at new optimization queries, avoiding repeated independent covariance estimation at every design. Active-learning acquisition functions allocate pilot samples to designs where covariance information is most uncertain and most relevant to the outer optimization objective. We demonstrate the method on a four-branch benchmark and a multi-fidelity Bayesian optimal experimental design application for sensor placement and measurement-time selection in a convection--diffusion system. The results show improved optimization performance relative to single-fidelity Monte Carlo and simpler pilot-sampling strategies.
\end{abstract}


\input{sections/1_intro}
\input{sections/2_background}

\input{sections/3_methods}
\input{sections/4_demos}
\input{sections/5_conclusions}

\section*{Acknowledgements}

This work is supported in part by the National Science Foundation Graduate Research Fellowship under Grant No. DGE 1841052, the Department of Navy awards N000142312735 and N000142512411 issued by the Office of Naval Research, and through computational resources and services provided by Advanced Research Computing at the University of Michigan, Ann Arbor.

\section*{Declaration of generative AI and AI-assisted technologies in the writing process}
During the preparation of this work, the authors used Gemini and ChatGPT to assist with language editing, organization, and clarity of presentation, as well as refactoring of the research code for public release. The authors reviewed and revised all AI-assisted text and take full responsibility for the content of the manuscript.

\bibliographystyle{unsrtnat_doi}
\bibliography{references}

\clearpage
\input{sections/6_appendix}

\end{document}

%% file: macros.tex
\renewcommand{\[}{\left[}

\newcommand{\vvvert}{|\kern-1pt|\kern-1pt|}

\newcommand{\EE}{\mathbb{E}}

\newcommand{\CA}{\mathcal{A}}

\newcommand{\CD}{\mathcal{D}}

\newcommand{\Var}{\mathbb{V}\text{ar}}
\newcommand{\Cov}{\mathbb{C}\text{ov}}

\newcommand{\design}{\xi}
\newcommand{\Design}{\Xi}

\floatname{algorithm}{Procedure}

%% file: sections/1_intro.tex
\section{Introduction}
\label{s: intro}

Uncertainty is pervasive in simulation-based engineering design and analysis. In many computational mechanics problems, a design must be chosen not from a single deterministic response of a numerical model, but from a statistic of a quantity of interest under uncertain material properties, boundary conditions, operating conditions, or forcing inputs. Such settings lead to optimization under uncertainty (OUU) problems, where each objective evaluation requires a forward uncertainty quantification (UQ) calculation. When the forward model is expensive, as is the case for many complex PDE-based physics models, this nested structure can make OUU computationally prohibitive: every design proposed by the outer optimization algorithm may require many evaluations of a high-fidelity model in order to estimate the objective. 
\Cref{fig:sf_ouu} illustrates this  structure. At iteration $k$, the optimizer proposes a design $\design_k$, and a forward UQ procedure returns a noisy estimate $\widehat{U}_0(\design_k)$ of the objective.

\tikzset{
    flowchart_style/.style={
        node distance=2cm and 1cm,
        box/.style={
            rectangle, 
            draw, 
            thick, 
            rounded corners,
            fill=gray!10,
            minimum height=1cm, 
            text width=2cm,
            align=center,     
            text centered,
            font=\sffamily\small
        },
        loop_arrow/.style={
            -Stealth,
            thick,
            rounded corners
        },
        label_style/.style={
            font=\small,
            midway
        }
    }
}

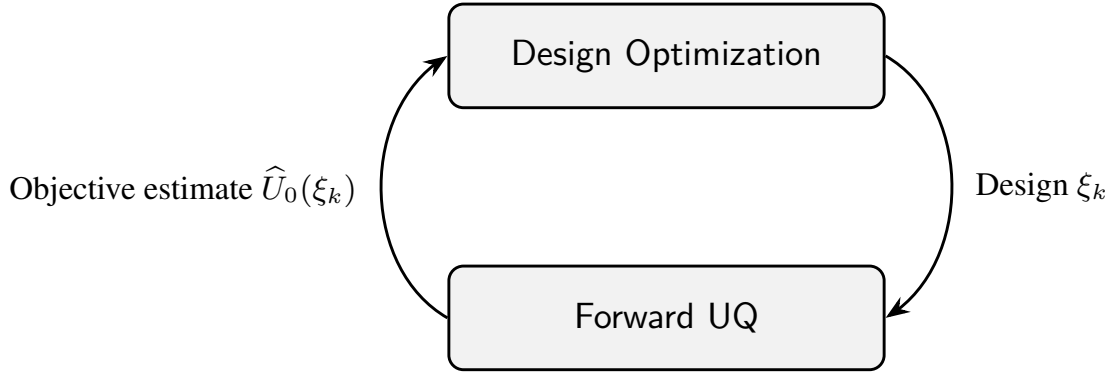
\begin{figure}[htbp]
  \centering
  \resizebox{0.9\linewidth}{!}{\begin{tikzpicture}[
    node distance=1.5cm and 1cm,
    box/.style={
        rectangle, 
        draw, 
        thick, 
        rounded corners,  
        fill=gray!10,
        minimum height=1.cm, 
        minimum width=4.2cm,
        text centered,
        font=\sffamily\normalsize
    },
    loop_arrow/.style={
        -Stealth, 
        thick,
        rounded corners  
    }
  ]
    \node (opt) [box] {Design Optimization};
    \node (uq) [box, below=of opt] {Forward UQ};

    \draw [loop_arrow] (opt.east) to [out=-30, in=30] 
        node[midway, right, xshift=1mm, font=\small] {Design $\design_k$} 
        (uq.east);
        
    \draw [loop_arrow] (uq.west) to [out=150, in=210] 
        node[midway, left, xshift=-1mm, font=\small] {Objective estimate $\widehat{U}_0(\design_k)$} 
        (opt.west);

  \end{tikzpicture}}
  \caption{Schematic of an iterative OUU procedure. At iteration $k$, the optimizer proposes a design $\design_k$, and forward UQ returns a noisy estimate $\widehat{U}_0(\design_k)$ of the objective value.}
  \label{fig:sf_ouu}
\end{figure}

A motivating example for this work is Bayesian optimal experimental design (OED)~\citep{chaloner_bayesian_1995,Ryan2016,Alexanderian2021,Rainforth2023,Strutz2024,Huan2024}. In Bayesian OED, one seeks experimental conditions that maximize an information-theoretic utility, most commonly the expected information gain (EIG)~\citep{lindley_measure_1956}.
For nonlinear models, the EIG generally does not have closed form and must be estimated using sampling-based methods. When these estimates require repeated evaluations of an expensive high-fidelity model, Bayesian OED becomes a challenging and representative OUU problem.

Multi-fidelity UQ methods provide one route to reducing the cost of the forward UQ step. These methods use an ensemble of cheaper low-fidelity models that are correlated with, but less accurate than, a high-fidelity model. Multi-fidelity estimators exploit the statistical dependence among model outputs, and can estimate high-fidelity model statistics at much lower cost than single-fidelity Monte Carlo (MC). Multilevel Monte Carlo (MLMC)~\citep{Giles2008} was an early and influential example of this idea, and has been applied to both black-box, gradient-free OUU~\citep{menhorn_multilevel_2024} and gradient-based stochastic optimization~\citep{hu_multi-level_2024,blanchet_unbiased_2019,dereich_multilevel_2011,alacaoglu_revisiting_2024}. However, MLMC is suited to model hierarchies with particular cost-accuracy relationships, such as those arising from mesh refinement or time-step refinement. In many mechanics applications, however, available model ensembles are not strict refinement hierarchies: they may combine different spatial or temporal discretizations, reduced-order models, simplified governing physics, empirical closures, or physically informed machine-learning surrogates.

Other multi-fidelity estimators relax these hierarchy requirements. 
Multi-fidelity Monte Carlo (MFMC)~\citep{peherstorfer_optimal_2016} combines estimators from multiple model fidelities in a nested control-variate structure. When model costs and the covariance matrix of the model outputs are known, MFMC can optimally reduce estimator variance while remaining unbiased with respect to the high-fidelity quantity of interest. Approximate control variate (ACV)~\citep{Gorodetsky2020,bomarito_optimization_2022} further generalizes this idea by allowing more flexible sample sharing across estimator components.
The multilevel best linear unbiased estimator (MLBLUE)~\citep{schaden_multilevel_2020}, based on linear-matrix theory, also provides substantial variance reduction, with connections to ACV explored through grouped ACV~\citep{gorodetsky_grouped_2024}. 
In the OED setting, an ACV-based multi-fidelity estimator for the EIG was introduced in~\cite{coons_multi-fidelity_2025}, where it achieved orders-of-magnitude variance reduction relative to single-fidelity MC.

A central practical issue for these methods is pilot sampling. To construct an efficient multi-fidelity estimator, one needs the model costs and the covariance matrix of model outputs. While model costs are often easy to estimate, the covariance matrix is generally unknown. A common approach is to run all models on an independent set of pilot samples, estimate the covariance matrix using the resulting model outputs, and then treat this estimate as exact when constructing the multi-fidelity estimator. This strategy can be expensive because it requires high-fidelity model evaluations before the main estimation task begins. Moreover, when only a limited number of pilot samples can be afforded, the estimated covariance matrix may be noisy, which can lead to suboptimal control-variate weights and sample allocations.

Recent work has begun to account for the cost and uncertainty of pilot sampling in one-off multi-fidelity estimation problems. One line of work develops adaptive explore-then-commit strategies using asymptotic analyses of sample covariance estimates~\citep{xu_bandit-learning_2022}, with later extensions to optimal sample allocations~\citep{dixon_optimally_2025}. Another approach uses Bayesian inference over the covariance matrix and an expected-loss criterion to decide when to terminate pilot sampling~\citep{coons_bayesian_2025}. 
These methods improve the treatment of pilot sampling for standalone forward UQ tasks, but they do not directly address the additional structure present in iterative OUU.

In an 
OUU loop, the covariance-estimation problem is not isolated to a single design. Instead, the optimizer queries a sequence of designs, and the model-output covariance may vary across the design domain. Existing multi-fidelity OUU methods have only partially addressed this issue. Early bi-fidelity OUU work~\citep{ng_multifidelity_2014} used both a low-fidelity model and information from the previous optimization iteration as control variates, but did not explicitly manage independent pilot sampling or the budget assigned to it. Other approaches rely on large pilot studies~\citep{adjei_multifidelity_2022, van_barel_mgopt_2021}, leave pilot sampling to future work~\citep{dixon_bistro_2025}, or do not explicitly address pilot sampling~\citep{de_bi-fidelity_2020,frikha_multi-level_2016}. The only multi-fidelity OUU methods that directly incorporate independent pilot sampling into the optimization loop are recent trust-region approaches~\citep{ha_adaptive_2025,ha_multi-fidelity_2025,menhorn_multilevel_2024}. 
These methods draw additional pilot samples at each optimization iteration until a trust-region accuracy condition is met~\citep{ha_adaptive_2025,ha_multi-fidelity_2025} or until the proposed sample allocation matches the number of pilot samples~\citep{menhorn_multilevel_2024}. However, these pilot samples are then reused in the corresponding multi-fidelity estimator, which induces bias, and the covariance information must be re-estimated from scratch at each iteration. 

The key observation motivating this work is that, in many OUU problems, covariance information should be shared across the design domain. If two nearby designs produce similar relationships among high- and low-fidelity model outputs, then their covariance matrices, and consequently their efficient multi-fidelity estimator settings, should also be similar. Conversely, if the covariance structure changes substantially over the design domain, then a single domain-averaged covariance matrix may produce inconsistent variance reduction. This behavior was observed in~\cite{coons_multi-fidelity_2025}, where a large pilot study was used to construct a design-averaged covariance matrix for multi-fidelity EIG estimation. That approach provided substantial gains, but it could not tailor the estimator to local covariance structure.

In this work, we formulate pilot sampling for multi-fidelity estimation in stochastic optimization as a covariance-emulation problem. Rather than estimating a covariance matrix independently at each optimization query, or estimating a single covariance matrix to use throughout the design domain, we learn a \emph{covariance surface over the design domain}. This surface shares information across pilot-sampled designs while providing design-specific covariance estimates at new query locations. The proposed framework has three components: (i) a positive-definite covariance parameterization and probabilistic emulator, (ii) an active-learning strategy for selecting pilot-design locations to learn the covariance surface, and (iii) a downstream stochastic optimization method that uses the resulting emulator to construct multi-fidelity objective estimates. Although our numerical demonstrations use Bayesian optimization and ACV-based estimators, the pilot-sampling framework is not tied to a particular stochastic optimization algorithm or to a particular control-variate-based multi-fidelity estimator.

The workflow proceeds in two stages. First, an offline pilot-sampling stage actively selects pilot-design locations, estimates local model-output covariances, and trains the covariance emulator. Second, after pilot sampling terminates, the trained emulator is passed to the downstream stochastic optimization stage, where it provides covariance information for constructing multi-fidelity estimators at each design queried by the optimizer. Thus, active learning is used to allocate pilot-sampling effort, while the outer optimizer is used to optimize the mean objective.

\begin{figure}[htbp]
  \centering

  \begin{subfigure}[b]{0.48\textwidth}
    \centering
    \resizebox{\linewidth}{!}{\begin{tikzpicture}[flowchart_style, node distance=1cm and 0.5cm]
      \node (opt) [box] {Design \\ Optimization};
      \node (midpoint) [coordinate, below=of opt, yshift=-0.5cm] {};
      \node (uq) [box, left=of midpoint, xshift=-.25cm] {Forward \\ UQ};
      \node (pilot) [box, right=of midpoint, xshift=.25cm] {Pilot \\ Sampling};

      \draw [loop_arrow] (opt.east) to [out=0, in=90] 
          node[label_style, right, xshift=1mm, pos=0.4] {$\design_k$} 
          (pilot.north);
      
      \draw [loop_arrow] (pilot.west) to
          node[label_style, above, yshift=1mm] {$\widehat{\Sigma}(\design_k)$}
          (uq.east);
          
      \draw [loop_arrow] (uq.north) to [out=90, in=180] 
          node[label_style, left, xshift=-1mm, pos=0.6] {$\widehat{U}_0(\design_k)$} 
          (opt.west);
    \end{tikzpicture}}
    \caption{Every-iteration pilot sampling}
    \label{subfig:iter-by-iter-loop}
  \end{subfigure}\hfill
  \begin{subfigure}[b]{0.48\textwidth}
    \centering
    \resizebox{\linewidth}{!}{\begin{tikzpicture}[flowchart_style, node distance=1cm and 1.2cm]
      \node (pilot) [box] {Pilot \\ Sampling};
      \node (emu) [box, below=of pilot] {Covariance \\ Averaging};
      \node (opt) [box, right=of pilot] {Design \\ Optimization};
      \node (uq) [box, below=of opt] {Forward \\ UQ};

      \draw [loop_arrow] (pilot.south) to 
          node[label_style, left, xshift=-1mm] {$\CD$} 
          (emu.north);

      \draw [loop_arrow] ([xshift=0.25cm]opt.south) to 
          node[label_style, right, xshift=1mm] {$\design_k$} 
          ([xshift=0.25cm]uq.north);

      \draw [loop_arrow] ([xshift=-0.25cm]uq.north) to 
          node[label_style, left, xshift=-1mm] {$\widehat{U}_0(\design_k)$} 
          ([xshift=-0.25cm]opt.south);
          
      \draw [loop_arrow] (emu.east) to
          node[label_style, above, yshift=1mm] {$\overline{\Sigma}$} 
          (uq.west);
    \end{tikzpicture}}
    \caption{Domain-averaged pilot sampling}
    \label{subfig:domain-avg-loop}
  \end{subfigure}

  \vspace{1.2em}

  \begin{subfigure}[b]{0.65\textwidth}
    \centering
    \resizebox{\linewidth}{!}{\begin{tikzpicture}[flowchart_style, node distance=1cm and 1.8cm]
      \node (pilot) [box] {Pilot \\ Sampling};
      \node (emu) [box, below=of pilot] {Covariance \\ Emulation};
      \node (opt) [box, right=of pilot] {Design \\ Optimization};
      \node (uq) [box, right=of emu] {Forward \\ UQ};

      \draw [loop_arrow] ([xshift=0.25cm]pilot.south) to 
          node[label_style, right, xshift=1mm] {$\CD$} 
          ([xshift=0.25cm]emu.north);

      \draw [loop_arrow] ([xshift=-0.25cm]emu.north) to 
          node[label_style, left, xshift=-1mm] {$\design_l$} 
          ([xshift=-0.25cm]pilot.south);

      \draw [loop_arrow] ([xshift=0.25cm]opt.south) to 
          node[label_style, right, xshift=1mm] {$\design_k$} 
          ([xshift=0.25cm]uq.north);

      \draw [loop_arrow] ([xshift=-0.25cm]uq.north) to 
          node[label_style, left, xshift=-1mm] {$\widehat{U}_0(\design_k)$} 
          ([xshift=-0.25cm]opt.south);
          
      \draw [loop_arrow] (emu.east) to
          node[label_style, above, yshift=1mm] {$p(\Sigma \vert \design_k, \CD)$} 
          (uq.west);
    \end{tikzpicture}}
    \caption{Proposed active-learning pilot sampling}
    \label{subfig:proposed-loop}
  \end{subfigure}

  \caption{Comparison of pilot-sampling strategies for iterative OUU. \textbf{(a)} A new covariance estimate is constructed independently at each optimization iteration, so no covariance information is shared across designs. \textbf{(b)} Pilot data $\CD$ are collected across the design domain and compressed into a single domain-averaged covariance matrix $\overline{\Sigma}$ for use in the multi-fidelity UQ estimator. \textbf{(c)} The pilot data are used to train a probabilistic covariance emulator, which both guides future pilot-sampling locations $\design_l$ and provides design-specific covariance information to the multi-fidelity UQ estimator at each queried design $\design_k$.}
  \label{fig:ouu_options}
\end{figure}
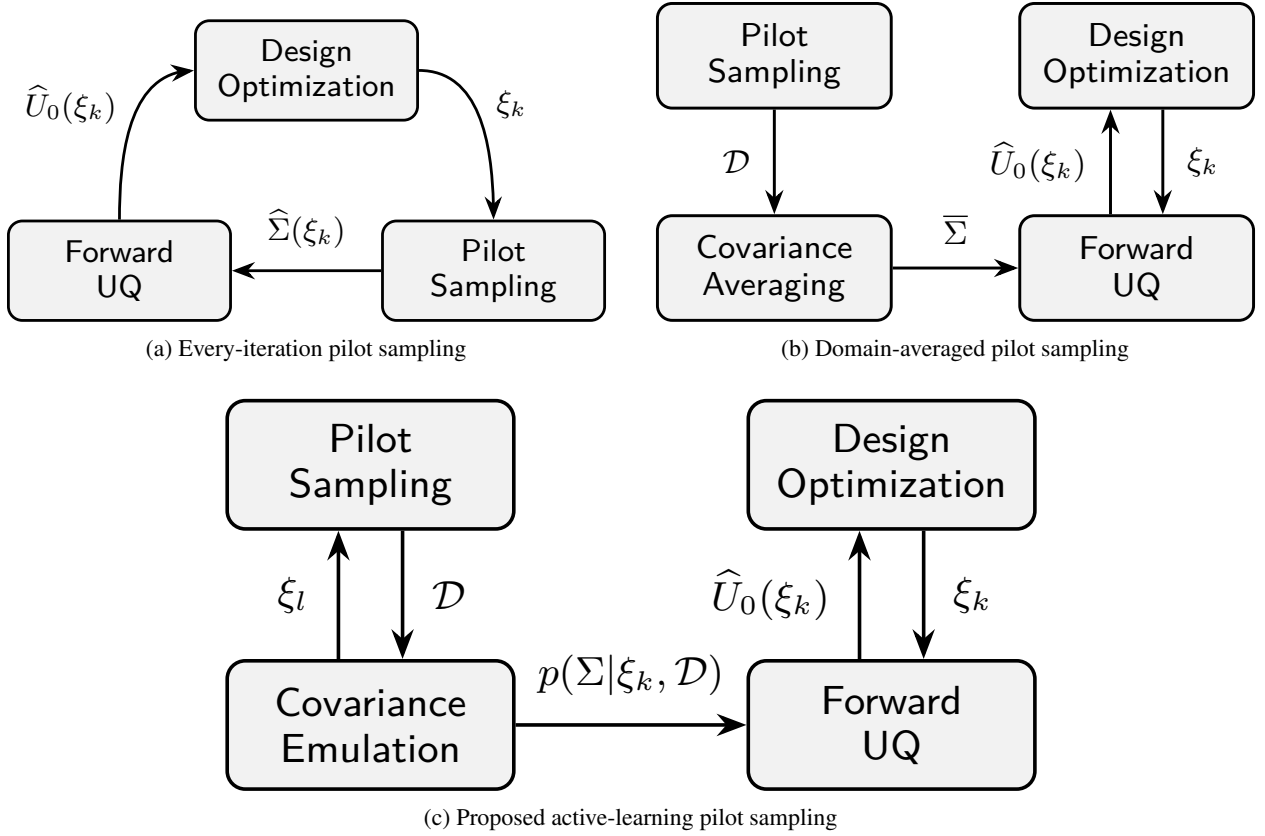

\Cref{fig:ouu_options} contrasts three pilot-sampling strategies for iterative OUU. In every-iteration pilot sampling (\cref{subfig:iter-by-iter-loop}), covariance is estimated independently at each queried design. This strategy provides local covariance information, but does not share information across designs. In domain-averaged pilot sampling (\cref{subfig:domain-avg-loop}), pilot data $\CD$ are collected across the design domain and compressed into a single covariance matrix $\overline{\Sigma}$. This strategy shares information globally, but cannot adapt to spatial variation in covariance structure. In the proposed approach (\cref{subfig:proposed-loop}), pilot data are used to train a probabilistic covariance emulator over the design domain. The emulator provides a posterior-predictive distribution $p(\Sigma | \design_k,\CD)$ at each optimization query $\design_k$ and guides future pilot-design locations $\design_l$ during the pilot-sampling stage. The resulting design-specific covariance estimates are then used to construct multi-fidelity estimators within the downstream stochastic optimization loop.

The purpose of the active-learning loop is to spend pilot-sampling budget where covariance information is most useful. In particular, the method aims to reduce posterior uncertainty in the covariance surface, with emphasis on regions expected to be important for the outer objective optimization. We do not solve the more general problem of optimally dividing a fixed computational budget between pilot sampling and objective optimization. Instead, we provide acquisition functions and covariance-emulation tools that can be used once a pilot-sampling budget or schedule has been chosen.

The main contributions of this work are as follows:
\begin{enumerate}
    \item We develop a Gaussian-process probabilistic emulator for a model-output covariance matrix over a design domain. The emulator uses a positive-definite-preserving parameterization, enabling covariance information to be predicted at new designs without requiring fresh independent pilot sampling at every optimization iteration.

    \item We introduce active-learning acquisition functions for pilot sampling across the design domain. These acquisition functions target posterior uncertainty in the covariance emulator and weight by the predicted objective value so that pilot sampling is concentrated in regions more relevant to the stochastic optimization problem.

    \item We demonstrate the proposed pilot-sampling framework on a multi-fidelity stochastic optimization problem and on a convection--diffusion Bayesian OED problem. The results show that covariance emulation can improve optimization performance relative to single-fidelity MC and simpler pilot-sampling strategies, even when the computational cost of pilot sampling is included.
\end{enumerate}
Our code is available at: \url{https://github.com/tcoonsUM/adaptive_pilot/tree/main}.

The remainder of the paper is organized as follows. \Cref{s: background} introduces the stochastic optimization setting and reviews the ACV estimator. \Cref{s: methodology} presents the covariance parameterization, the probabilistic covariance emulator, and the active-learning acquisition functions used for pilot sampling. \Cref{s: results} demonstrates the method on a benchmark problem and on a Bayesian OED application for contaminant source inversion. \Cref{s: conclusions} summarizes the main findings, limitations, and directions for future work.

%% file: sections/2_background.tex
\section{Background}
\label{s: background}
We first introduce the expected-objective stochastic optimization problem considered in this work, then review the ACV estimator used for multi-fidelity estimation.

\subsection{Stochastic optimization}\label{ss:stoch-opt}

Consider a high-fidelity predictive model $f_0(\Theta,\design)$ with uncertain input $\Theta\footnote{Where appropriate, we use capital letters to denote random variables and lowercase letters to denote their realizations and other non-random quantities.} \in \mathbb{R}^{n_{\theta}}$ and deterministic design variable $\design \in \Design$. We define the corresponding random quantity of interest as
\begin{align}
    Q_0(\design) = f_0(\Theta,\design).
\end{align}
Since $Q_0(\design)$ is random, directly optimizing the model response over $\Design$ is generally not meaningful without first specifying how uncertainty should be aggregated. OUU addresses this issue by replacing the random response with a deterministic risk measure that summarizes performance over the distribution of uncertain inputs. Different choices emphasize different aspects of the response distribution, including mean, variance, quantiles, or superquantiles.

In this work, we focus on the common risk-neutral setting, where performance is measured by the expected response,
\begin{align}
    U_0(\design) := \EE\left[Q_0(\design)\right].
    \label{eqn:g}
\end{align}
The resulting expected-objective OUU problem takes the form of a stochastic optimization problem,
\begin{align}
    \design^{\ast} \in \arg \max_{\design \in \Design} U_0(\design).
    \label{eqn:stoch_opt}
\end{align}
This usage is distinct from stochastic programming in the classical sense, which often refers to optimization problems with a linear objective and recourse under uncertainty~\citep{giunta_perspectives_2004,birge_introduction_2011}.

Solving \eqref{eqn:stoch_opt} is challenging because evaluating $U_0(\design)$ requires computing the expectation in \eqref{eqn:g} at each design queried by the optimization algorithm. A standard approach is to use the $N$-sample MC estimator
\begin{align}
    U_0(\design) \approx \widehat{U}_0(\design; \boldsymbol{\theta})
    := \frac{1}{N}\sum_{j=1}^{N} f_0\left(\theta^{(j)},\design\right),
    \label{eqn:mc}
\end{align}
where $\boldsymbol{\theta}=\{\theta^{(j)}\}_{j=1}^{N}$ and $\theta^{(j)} \sim p(\Theta)$ are independent and identically distributed (i.i.d.) samples. This estimator is unbiased, with variance
\begin{align}
    \Var\left[\widehat{U}_0(\design; \boldsymbol{\Theta})\right]
    = \frac{1}{N}\Var\left[f_0(\Theta,\design)\right].
\end{align}
Thus, increasing $N$ reduces objective-estimation noise but increases the cost of each optimization query.

This noise-cost tradeoff is central to stochastic optimization. Existing methods differ primarily in how they allocate samples during the optimization process. Sample average approximation (SAA) fixes a common sample set and optimizes the resulting deterministic MC approximation, allowing deterministic optimization algorithms to be used but potentially introducing solution error unless the sample size is large. Stochastic approximation (SA) methods, such as stochastic gradient descent,
use small numbers of samples per iteration to construct noisy objective or gradient information; these methods can be highly efficient when gradients are available, but global convergence typically requires assumptions such as convexity or smoothness. Response surface methodology (RSM) instead builds a surrogate model of the objective surface and uses that surrogate to decide where to sample next.

Bayesian optimization (BO) is a common RSM approach for expensive black-box objectives~\citep{movckus1975bayesian,jones1998efficient,shahriari2015taking,frazier2018tutorial}.
BO typically uses a probabilistic surrogate, often a Gaussian process, together with an acquisition function that balances exploration of uncertain regions and exploitation of promising regions. BO is therefore well suited to the derivative-free, nonconvex settings that arise in Bayesian OED. However, its performance depends on the surrogate's ability to learn the objective surface, which can be difficult in high-dimensional design spaces or under limited data~\citep{doumont_we_2026}.

In the numerical demonstrations of this paper, we use BO to optimize noisy estimates of the objective, with ACV-based multi-fidelity estimators replacing single-fidelity MC evaluations. 
The pilot-sampling methodology developed in this paper is not specific to BO. Its role is to provide design-dependent model covariance estimates that can be used to construct performant multi-fidelity estimators in lieu of single-fidelity MC estimators within any compatible stochastic optimization algorithm.

\subsection{Approximate control variates}

The number of high-fidelity model evaluations required by the MC estimator in \eqref{eqn:mc} can be prohibitively large. Multi-fidelity estimation methods reduce this cost by exploiting cheaper low-fidelity models that are correlated with the high-fidelity model. In this subsection, we review the ACV estimator~\citep{Gorodetsky2020,bomarito_optimization_2022} in the one-off estimation setting, suppressing the explicit dependence on $\design$ for notational simplicity. Throughout, $\Theta$ denotes a single random input, while
$\boldsymbol{\Theta}$ denotes a finite random sample set.

Let $Q_0=f_0(\Theta)$ denote the high-fidelity quantity of interest, and let $Q_m=f_m(\Theta)$, $m=1,\ldots,M-1$, denote auxiliary low-fidelity quantities. The ACV estimator uses MC estimates of these auxiliary quantities as control variates:
\begin{align}
    \widetilde{Q}(\boldsymbol{\theta};\alpha,\CA)
    &=
    \widehat{Q}_0(\boldsymbol{\theta}_0)
    +
    \sum_{m=1}^{M-1}
    \alpha_m
    \left(
        \widehat{Q}_m(\boldsymbol{\theta}^{\ast}_m)
        -
        \widehat{Q}_m(\boldsymbol{\theta}_m)
    \right),
    \label{eqn:acv_formula}
\end{align}
where $\alpha=[\alpha_1,\ldots,\alpha_{M-1}]^\top$ are the control-variate weights. 
Here, $\boldsymbol{\theta}$ is a realization of the random sample set
$\boldsymbol{\Theta}$. The sample allocation $\CA$ determines how
$\boldsymbol{\Theta}$ is partitioned into the subsets
$\boldsymbol{\Theta}_0$, $\boldsymbol{\Theta}_m$, and
$\boldsymbol{\Theta}^{\ast}_m$, with corresponding realizations
$\boldsymbol{\theta}_0$, $\boldsymbol{\theta}_m$, and
$\boldsymbol{\theta}^{\ast}_m$.
The difference terms have zero mean, so the corresponding random estimator
$\widetilde{Q}(\boldsymbol{\Theta};\alpha,\CA)$ is unbiased with respect to
$Q_0$ whenever the component MC estimators are unbiased and the control-variate
weights are independent of the samples used in the final estimator.

For compactness, define
\begin{align}
    \Delta_m
    &:=
    \widehat{Q}_m(\boldsymbol{\Theta}^{\ast}_m)
    -
    \widehat{Q}_m(\boldsymbol{\Theta}_m),
    \qquad m=1,\ldots,M-1,
    \\
    \Delta
    &:=
    [\Delta_1,\ldots,\Delta_{M-1}]^\top.
\end{align}
Then \eqref{eqn:acv_formula} can be written as
\begin{align}
    \widetilde{Q}(\boldsymbol{\Theta};\alpha,\CA)
    =
    \widehat{Q}_0(\boldsymbol{\Theta}_0)
    +
    \alpha^\top \Delta.
    \label{eqn:vectorized-acv}
\end{align}
The corresponding estimator variance is
\begin{align}
    \Var\left[
        \widetilde{Q}(\boldsymbol{\Theta};\alpha,\CA)
    \right]
    &=
    \Var\left[
        \widehat{Q}_0(\boldsymbol{\Theta}_0)
    \right]
    +
    \alpha^\top
    \Cov\left[\Delta,\Delta\right]
    \alpha
    +
    2\alpha^\top
    \Cov\left[
        \Delta,\widehat{Q}_0(\boldsymbol{\Theta}_0)
    \right].
    \label{eqn:acv-variance}
\end{align}
For a fixed sample allocation $\CA$, minimizing \eqref{eqn:acv-variance} with respect to $\alpha$ gives the optimal weights
\begin{align}
    \alpha^{\ast}(\CA)
    =
    -
    \Cov\left[\Delta,\Delta\right]^{-1}
    \Cov\left[
        \Delta,\widehat{Q}_0(\boldsymbol{\Theta}_0)
    \right].
    \label{eqn:alpha-star-acv}
\end{align}
Substituting \eqref{eqn:alpha-star-acv} into \eqref{eqn:acv-variance} yields the minimized estimator variance~\citep{Gorodetsky2020},
\begin{align}
    \Var\left[
        \widetilde{Q}^{\alpha^{\ast}}(\CA)
    \right]
    &=
    \Var\left[
        \widehat{Q}_0(\boldsymbol{\Theta}_0)
    \right]
    -
    \Cov\left[
        \Delta,\widehat{Q}_0(\boldsymbol{\Theta}_0)
    \right]^\top
    \Cov\left[\Delta,\Delta\right]^{-1}
    \Cov\left[
        \Delta,\widehat{Q}_0(\boldsymbol{\Theta}_0)
    \right].
    \label{eqn:acv-variance-opt}
\end{align}

The sample allocation $\CA$ determines how computational effort is distributed across the model ensemble. The set of all possible allocations can be large, so practical ACV methods restrict $\CA$ to a prescribed allocation family. Classical MLMC~\citep{giles_multilevel_2015} and MFMC~\citep{peherstorfer_optimal_2016} can be interpreted as particular allocation strategies (also referred to as ``estimators'' in some literature) with additional structure and analytical solutions, 
while more flexible ACV estimators often require numerical optimization over the allocation variables. Given a set $\mathbb{A}$ of admissible allocations, efficient tools have been developed to approximately solve allocation problems of the form~\citep{bomarito_multi_2020,jakeman_pyapprox_2023}
\begin{align}
    \min_{\CA \in \mathbb{A}}
    &\quad
    \Var\left[
        \widetilde{Q}(\boldsymbol{\Theta};\alpha^{\ast}(\CA),\CA)
    \right]
    \label{eqn:mxmcpy}
    \\
    \mathrm{subject \,\, to}
    &\quad
    \mathcal{W}(w,\CA) \leq B,
    \label{eqn:mxmcpy_constraint}
\end{align}
where $B$ is the total computational budget, $w=[w_0,\ldots,w_{M-1}]^\top$ are the per-sample model costs, and $\mathcal{W}(w,\CA)$ denotes the total estimator cost.

Both the optimal weights in \eqref{eqn:alpha-star-acv} and the allocation problem in \eqref{eqn:mxmcpy} depend on the model-output covariance matrix
\begin{align}
    \Sigma
    =
    \Cov\left[
        f_0(\Theta),\ldots,f_{M-1}(\Theta)
    \right].
    \label{eqn:cov_defn}
\end{align}
The costs $w$ are often known or inexpensive to estimate, but $\Sigma$ is generally unknown. A standard approach is to estimate it from an independent pilot sample set. Given $N_{\mathrm{pilot}}$ pilot samples, the unbiased sample covariance estimator is
\begin{align}
    \widehat{\Sigma}
    =
    \frac{1}{N_{\mathrm{pilot}}-1}
    \sum_{j=1}^{N_{\mathrm{pilot}}}
    \left(
        f^{(j)}-\overline{f}
    \right)
    \left(
        f^{(j)}-\overline{f}
    \right)^\top,
    \label{eqn:cov}
\end{align}
where 
\begin{align}
    f^{(j)}
    &=
    \left[
        f_0(\theta^{(j)}),\ldots,f_{M-1}(\theta^{(j)})
    \right]^\top,
    \\
    \overline{f}
    &=
    \left[
        \overline{f}_0,\ldots,\overline{f}_{M-1}
    \right]^\top,
    \qquad
    \overline{f}_m
    =
    \frac{1}{N_{\mathrm{pilot}}}
    \sum_{j=1}^{N_{\mathrm{pilot}}}
    f_m(\theta^{(j)}).
\end{align}
Accurate covariance estimation is therefore essential for designing efficient multi-fidelity estimators under limited computational budgets. Recent work has developed pilot-sampling strategies for one-off estimation, including adaptive sample-allocation approaches~\citep{xu_bandit-learning_2022,dixon_optimally_2025} and Bayesian covariance-inference methods that use posterior summaries of $\Sigma$ in place of the empirical covariance estimate~\citep{coons_bayesian_2025}. In this work, we extend the pilot-sampling question from one-off estimation to iterative stochastic optimization, where covariance information must be learned and reused across a design domain.

%% file: sections/3_methods.tex
\section{Methods}
\label{s: methodology}

We now develop the pilot-sampling stage of the proposed framework, in which a covariance emulator is constructed and active-learning acquisition functions are used to select pilot-design locations.

\subsection{Probabilistic emulation of the model covariance matrix}\label{ss:gp-gamma}

For multi-fidelity estimation within an outer optimization loop, the relevant model-output covariance generally depends on the design. We therefore seek an emulator for the covariance-matrix function
\begin{align}
    \Sigma(\design)
    =
    \Cov\left[
        f_0(\Theta,\design),\ldots,f_{M-1}(\Theta,\design)
    \right],
    \qquad
    \design \in \Design.
    \label{eqn:cov_function_defn}
\end{align}
Modeling \eqref{eqn:cov_function_defn} directly is difficult because each covariance matrix must satisfy symmetry and positive-semidefiniteness constraints. 
Covariance interpolation approaches using matrix-valued stochastic processes, such as Wishart processes and their generalized variants \citep{WilsonGWP}, directly emulate the full PD matrix. Wishart processes provide embedded UQ and fast prediction but require MCMC for hyperparameter tuning, making them numerically cumbersome. Other methods interpolate parameterizations of the covariance matrix, which we also adopt in this work.

We assume the relevant covariance matrices are positive definite and use a bijective transformation from $\Sigma(\design)$ to an unconstrained vector of real-valued parameters. These transformed parameters can then be modeled with standard regression tools. Several such parameterizations have been developed, including Cholesky decompositions~\citep{pinheiro_unconstrained_1996}, Givens-rotation parameterizations~\citep{rapisarda_parameterizing_2007,rebonato_most_2011}, and matrix-logarithm parameterizations~\citep{archakov_new_2021,higham_11_2008,chiu_matrix-logarithmic_1996}. Motivated by empirical comparisons in~\cite{bucci_comparing_2022}, we adopt a matrix-logarithm parameterization of the \emph{correlation matrix}, which we refer to as the $\gamma$-transform.

\paragraph{Bijective parameterization of the covariance matrix}\label{par:bij-param}

\begin{figure}[hbtp]
\centering
\resizebox{\linewidth}{!}{\begin{tikzpicture}[
    node distance=.5cm and 1.2cm, 
    box/.style={
        rectangle, 
        draw, 
        thick, 
        rounded corners,  
        fill=gray!10,
        minimum height=1.3cm, 
        minimum width=2.8cm,
        text centered,
        font=\sffamily\large
    },
    loop_arrow/.style={
        Stealth-Stealth, 
        thick,
        rounded corners
    }
]

    \node[box] (cov) {
        $\Sigma(\xi) \in \mathbb{S}_{++}^{M \times M}$
    };
    \node[above=0.2cm of cov, font=\sffamily\bfseries] {Covariance};

    \node[box, above right=0.3cm and 1.2cm of cov] (corr) {
        $R(\xi) \in \mathbb{S}_{++}^{M \times M}$
    };
    \node[above=0.2cm of corr, font=\sffamily\bfseries] {Correlation};

    \node[box, below right=0.3cm and 1.2cm of cov] (std) {
        $\sigma(\xi) \in \mathbb{R}_{+}^{M}$
    };
    \node[above=0.2cm of std, font=\sffamily\bfseries] {Standard Deviations};

    \node[box, right=1.2cm of corr] (gamma) {
        $\gamma(\xi) \in \mathbb{R}^{M(M-1)/2}$
    };

    \node[box, right=1.2cm of std] (logsig) {
        $\log( \sigma(\xi)) \in \mathbb{R}^{M}$
    };

    \node[font=\sffamily\bfseries, align=center] at ($(gamma.north)!0.5!(logsig.south) $) {Unbounded\\Parameters\\$\mathbb{R}^{M(M+1)/2}$};

    \draw[loop_arrow] (cov.east) -- (corr.west);
    \draw[loop_arrow] (cov.east) -- (std.west);
    \draw[loop_arrow] (corr) -- node[above, font=\small, yshift=1mm] {} (gamma);
    \draw[loop_arrow] (std) -- node[above, font=\small, yshift=1mm] {} (logsig);

\end{tikzpicture}}
\caption{Bijective parameterization of the covariance matrix. The covariance matrix $\Sigma(\design)$ is decomposed into a correlation matrix $R(\design)$ and standard deviations $\sigma(\design)$. The correlation matrix is mapped to unconstrained $\gamma$-parameters through the matrix-logarithm transform, while the standard deviations are mapped to unconstrained log-standard-deviation parameters.}
\label{fig:cov_parameterization}
\end{figure}
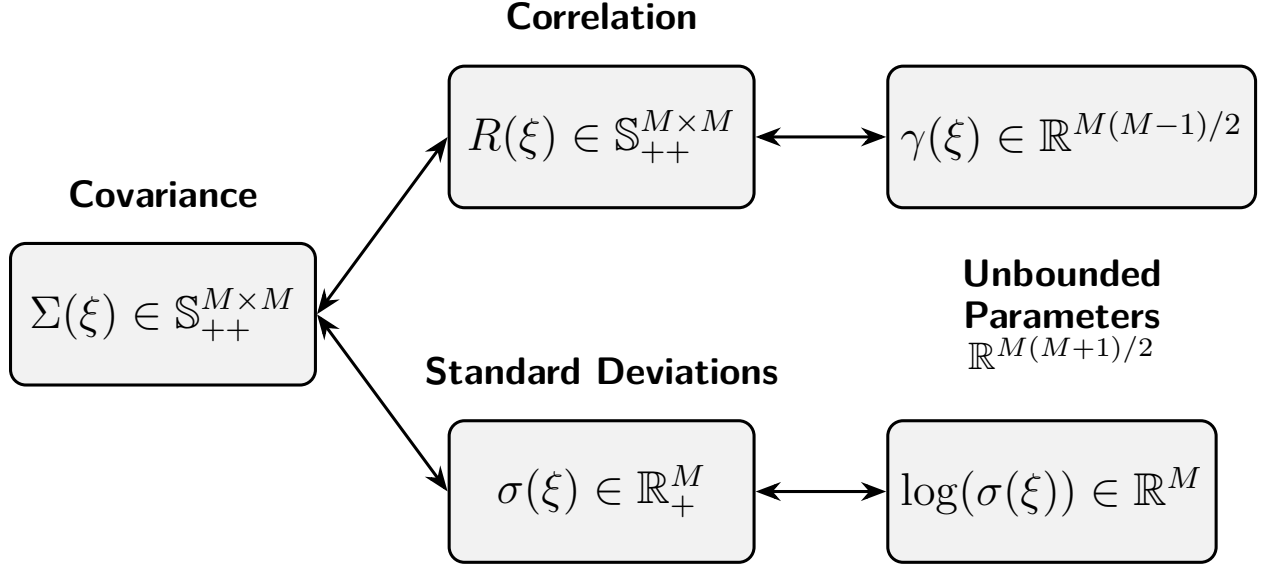

We first decompose the covariance matrix into standard deviations and correlations,
\begin{align}
    \Sigma(\design)
    =
    D(\design) R(\design) D(\design),
    \label{eqn:cov_to_corr_sigma}
\end{align}
where $D(\design)$ is diagonal with entries $\sigma(\design) \in \mathbb{R}_{+}^{M}$, and $R(\design)$ is the corresponding correlation matrix. We model the standard deviations through their logarithms,
\begin{align}
    \ell(\design)
    :=
    \log \sigma(\design)
    =
    \log\left(\diag(D(\design))\right).
    \label{eqn:log_std_defn}
\end{align}
Let $\Gamma$ denote the forward $\gamma$-transform and let $\mathrm{vecl}(A)$ denote the vectorization of the strict lower-triangular elements of a square matrix $A$. The $\gamma$-transform is defined by
\begin{align}
    \gamma(\design)
    =
    \Gamma(R(\design))
    =
    \mathrm{vecl}\left(\log(R(\design))\right),
    \label{eqn:gamma_mapping}
\end{align}
where $\log(R(\design))$ is the matrix logarithm. The vector $\gamma(\design)$ contains
\begin{align}
    N_{\gamma}
    =
    \frac{M(M-1)}{2}
    \label{eqn:n_gamma_defn}
\end{align}
correlation parameters. Altogether, $\gamma(\design)$ and $\ell(\design)$ provide $M(M+1)/2$ unconstrained parameters that fully determine $\Sigma(\design)$. The mapping is illustrated in \cref{fig:cov_parameterization}. We use the numerical implementations from~\cite{archakov_new_2021} for both the forward and inverse $\gamma$-transforms.

\paragraph{Gaussian process emulator}

After the covariance matrix has been mapped to unconstrained parameters, we model each transformed component as a function of the design. We use independent Gaussian process (GP) emulators~\citep{rasmussen_gaussian_2008} for each correlation parameter and each log standard deviation. This independent-emulator construction is a pragmatic modeling choice: it provides tractable posterior predictions and uncertainty estimates under limited pilot data, while preserving positive definiteness after mapping back to the covariance space. More structured multi-output emulators could also be used, but we do not pursue them here.

For $i=1,\ldots,N_{\gamma}$ and $m=0,\ldots,M-1$, we place independent GP priors
\begin{align}
    \gamma_i(\design)
    &\sim
    \mathcal{GP}\left(
        \mu_{\gamma_i}(\design),
        k_{\gamma_i}(\design,\design')
    \right),
    \\
    \ell_m(\design)
    =
    \log \sigma_m(\design)
    &\sim
    \mathcal{GP}\left(
        \mu_{\ell_m}(\design),
        k_{\ell_m}(\design,\design')
    \right),
    \label{eqn:gp-gamma}
\end{align}
where the $\mu$ terms and $k$ terms denote the corresponding GP mean and covariance functions. 
In the numerical demonstrations, we use constant mean functions and radial basis function kernels, with hyperparameters estimated by maximum likelihood. The use of Gaussian likelihoods for the transformed covariance observations is also motivated by evidence that the sampling distribution of the $\gamma$-parameters is well approximated by a Gaussian distribution for sufficiently large sample sizes~\citep{archakov_new_2021}.

Suppose pilot sampling has been performed so far at design locations $\{\design_j\}_{j=1}^{N_{\CD}}$, with $\design_j \in \Design \subseteq \mathbb{R}^{d_{\design}}$. Let
\begin{align}
    X
    &:=
    [\design_1,\ldots,\design_{N_{\CD}}]^\top
    \in
    \mathbb{R}^{N_{\CD} \times d_{\design}}
\end{align}
denote the design matrix. Each pilot-sampling batch (at each design) provides a sample covariance estimate, which is transformed into noisy observations of the unconstrained parameters. We collect these observations as
\begin{align}
    G
    &:=
    [G_1,\ldots,G_{N_{\gamma}}]
    \in
    \mathbb{R}^{N_{\CD} \times N_{\gamma}},
    \qquad
    G_i
    :=
    [
        \widehat{\gamma}_{i,1},\ldots,\widehat{\gamma}_{i,N_{\CD}}
    ]^\top,
    \\
    S
    &:=
    [S_0,\ldots,S_{M-1}]
    \in
    \mathbb{R}^{N_{\CD} \times M},
    \qquad
    S_m
    :=
    [
        \log \widehat{\sigma}_{m,1},\ldots,\log \widehat{\sigma}_{m,N_{\CD}}
    ]^\top.
\end{align}
The full pilot dataset is denoted by
\begin{align}
    {\CD}
    :=
    \{X,G,S\}.
\end{align}

The observation model for the transformed pilot estimates is
\begin{align}
    \widehat{\gamma}_{i,j}
    &=
    \gamma_i(\design_j)
    +
    \epsilon_{i,j},
    \qquad
    \epsilon_{i,j}
    \sim
    \mathcal{N}(0,\tau_{i,j}^2),
    \\
    \log \widehat{\sigma}_{m,j}
    &=
    \ell_m(\design_j)
    +
    \zeta_{m,j},
    \qquad
    \zeta_{m,j}
    \sim
    \mathcal{N}(0,\eta_{m,j}^2),
    \label{eqn:gp-likelihood}
\end{align}
for $i=1,\ldots,N_{\gamma}$, $m=0,\ldots,M-1$, and $j=1,\ldots,N_{\CD}$. The observation noise arises because each transformed quantity is computed from a finite number of pilot samples. To estimate the noise variances $\tau_{i,j}^2$ and $\eta_{m,j}^2$, we use pushforward samples from the sampling distribution of the sample covariance matrix. If the $j$th pilot batch contains $N_{\mathrm{pilot},j}$ samples, then, under Gaussian sampling approximation,
\begin{align}
    \widehat{\Sigma}_j
    \sim
    \frac{1}{N_{\mathrm{pilot},j}-1}
    \mathcal{W}_M
    \left(
        \Sigma_j,
        N_{\mathrm{pilot},j}-1
    \right),
    \label{eqn:wishart_sampling_dist}
\end{align}
where $\mathcal{W}_M$ denotes the $M$-dimensional Wishart distribution and $\Sigma_j=\Sigma(\design_j)$. Samples from \eqref{eqn:wishart_sampling_dist} are propagated through the $\gamma$-transform and the logarithm of the standard deviations to estimate the corresponding observation-noise variances, as described in Section~3.2.2 of~\cite{coons_bayesian_2025}.

For a new prediction design $\design'$, the GP posterior predictive distribution for each transformed component is Gaussian:
\begin{align}
    p(\gamma_i' | \design',{\CD})
    &=
    \mathcal{N}
    \left(\gamma_i'; 
        \mu_{\gamma_i'|{\CD}},
        K_{\gamma_i'|{\CD}}
    \right),
    \\
    p(\ell_m' | \design',{\CD})
    &=
    \mathcal{N}
    \left(\ell_m'; 
        \mu_{\ell_m'|{\CD}},
        K_{\ell_m'|{\CD}}
    \right).
    \label{eqn:gp-post-pred}
\end{align}
For the $\gamma_i$ emulator, define
\begin{align}
    \mathcal{E}_i
    :=
    \diag(\tau_{i,1}^2,\ldots,\tau_{i,N_{\CD}}^2).
\end{align}
Then the posterior predictive mean and covariance are
\begin{align}
    \mu_{\gamma_i'|{\CD}}
    &=
    \mu_{\gamma_i}(\design')
    +
    k_{\gamma_i}(\design',X)^\top
    \left(
        K_{\gamma_i}(X,X)
        +
        \mathcal{E}_i
    \right)^{-1}
    \left(
        G_i
        -
        \mu_{\gamma_i}(X)
    \right),
    \label{eqn:gp-post-pred-mean}
    \\
    K_{\gamma_i'|{\CD}}
    &=
    k_{\gamma_i}(\design',\design')
    -
    k_{\gamma_i}(\design',X)^\top
    \left(
        K_{\gamma_i}(X,X)
        +
        \mathcal{E}_i
    \right)^{-1}
    k_{\gamma_i}(\design',X).
    \label{eqn:gp-post-pred-cov}
\end{align}
The corresponding posterior predictive mean and covariance for each log-standard-deviation emulator are obtained by replacing $G_i$, $\mu_{\gamma_i}$, $k_{\gamma_i}$, $K_{\gamma_i}$, and $\mathcal{E}_i$ with $S_m$, $\mu_{\ell_m}$, $k_{\ell_m}$, $K_{\ell_m}$, and
\begin{align}
    \mathcal{H}_m
    :=
    \diag(\eta_{m,1}^2,\ldots,\eta_{m,N_{\CD}}^2),
\end{align}
respectively.

This construction defines a probabilistic emulator over unconstrained transformed covariance parameters. Posterior samples of $\gamma(\design')$ and $\ell(\design')$ can be mapped through the inverse parameterization to obtain posterior samples of $R(\design')$ and $\Sigma(\design')$. Similarly, posterior means of the transformed parameters can be mapped back to a positive-definite covariance estimate for use in the multi-fidelity estimator.

\paragraph{Leveraging the covariance emulator for multi-fidelity stochastic optimization}

\begin{sloppypar}
Once the covariance emulator has been trained, the outer optimization algorithm can query new designs without requiring fresh pilot sampling at each design. At a queried design $\design_k$, the emulator provides design-specific covariance information that can be used to construct ACV estimators with covariance-dependent weights and sample allocations. In the numerical demonstrations, we use the posterior mean predictions of the $\gamma$- and $\ell$-GPs, mapped back through the inverse parameterization in \cref{par:bij-param}, to determine the ACV estimator hyperparameters.    
\end{sloppypar}

The emulator also provides uncertainty estimates. Some stochastic optimization algorithms, including noisy BO with heteroscedastic observation models, can use an estimate of the objective-estimation variance at each queried design. A natural estimate is the ACV variance in \eqref{eqn:acv-variance-opt}, evaluated using the posterior mean covariance prediction. However, this plug-in estimate can be overconfident in regions where the covariance emulator itself is uncertain. A fully Bayesian correction would propagate posterior samples of $\Sigma(\design_k)$ through the ACV hyperparameter optimization and variance formula. In particular, each posterior sample may induce a different optimal sample allocation, and propagating the posterior samples through different sample allocations would not be computationally feasible. To avoid this cost, we use a heuristic in the demonstrations: we inflate the ACV variance estimate by accounting for posterior uncertainty in the high-fidelity variance. Specifically, we use an upper posterior quantile (e.g., two standard deviations above the mean) of the high-fidelity log-standard-deviation emulator, rather than its posterior mean, when evaluating the ACV variance. This heuristic does not fully propagate covariance-emulator uncertainty through the ACV construction, but it guards against underestimating the objective-estimation noise in regions where the high-fidelity variance is uncertain.

\subsection{Active learning for pilot sampling}
\label{ss:al-gamma}

The covariance emulator provides posterior predictive uncertainty over the transformed covariance parameters, making active learning a natural mechanism for selecting pilot-design locations. The active-learning procedure described here is used only during the pilot-sampling stage: its purpose is to decide where to collect covariance information, not to select the designs queried later by the downstream stochastic optimization algorithm.

At pilot-sampling iteration $l$, the next pilot-design location is selected by maximizing an acquisition function over the design domain,
\begin{align}
    \design_l
    \in
    \arg\max_{\design\in\Design}
    a_l(\design;\mathcal{D}).
    \label{eqn:al-opt}
\end{align}
The acquisition function encodes how pilot-sampling effort should be allocated. In this work, we aim to learn the covariance surface efficiently while emphasizing regions that are likely to be important for the subsequent objective-optimization stage.

We consider two acquisition functions. The first, denoted $a_{\mathrm{us}}$, is an uncertainty-sampling acquisition function that targets posterior uncertainty in the covariance emulator. The second, denoted $a_{\mathrm{mw}}$, is a mean-weighted uncertainty-sampling acquisition function that also targets covariance uncertainty, but gives greater priority to designs that are predicted to have larger objective values.

Using the posterior predictive covariance notation from \eqref{eqn:gp-post-pred-cov}, define the total transformed-parameter uncertainty at design $\design$ as
\begin{align}
    s_{\Sigma}(\design;\mathcal{D})
    &:=
    \sum_{i=1}^{N_{\gamma}}
    \sqrt{
        \det\left[
            K_{\gamma_i|\mathcal{D}}(\design)
        \right]
    }
    +
    \sum_{m=0}^{M-1}
    \sqrt{
        \det\left[
            K_{\ell_m|\mathcal{D}}(\design)
        \right]
    }.
    \label{eqn:cov-uncertainty-score}
\end{align}
Here, $K_{\gamma_i|\mathcal{D}}(\design)$ is the posterior predictive covariance for the $i$th $\gamma$-parameter and $K_{\ell_m|\mathcal{D}}(\design)$ is the posterior predictive covariance for the $m$th log standard deviation. When a single design, rather than a batch of designs, is queried at each pilot-sampling iteration, these predictive covariances are scalar variances, so \eqref{eqn:cov-uncertainty-score} reduces to the sum of posterior predictive standard deviations over all transformed covariance parameters.

The uncertainty-sampling acquisition function is
\begin{align}
    a_{\mathrm{us}}(\design;\mathcal{D})
    &:=
    s_{\Sigma}(\design;\mathcal{D}).
    \label{eqn:us-acquisition}
\end{align}
This acquisition function prioritizes designs where the covariance emulator is most uncertain, independent of whether those designs are likely to be important for the outer optimization problem. It is therefore an exploratory criterion for learning the covariance surface globally.

For stochastic optimization, however, covariance accuracy is most valuable in regions that are likely to be queried repeatedly by the outer optimizer or that are expected to contain high objective values. To incorporate this information, we introduce an auxiliary emulator for the mean objective $U_0(\design)$. This emulator can be trained using the high-fidelity subset of the pilot data, so no additional forward model evaluations are needed. In RSM methods such as BO, the same data can also be used to initialize the objective surrogate. Let $U_0^{\mathrm{sim}}(\design;\mathcal{D})$ denote the predicted objective value after observing pilot data $\mathcal{D}$, where this quantity may be a posterior predictive mean for a probabilistic emulator or a point prediction for a deterministic emulator.

We convert the objective prediction into an importance weight before using it multiplicatively with the covariance-uncertainty score. Let $\omega_U(\design;\mathcal{D})$ denote this weight, chosen so that when positive, designs with larger predicted objective values receive larger weights. The mean-weighted uncertainty-sampling acquisition function is then
\begin{align}
    a_{\mathrm{mw}}(\design;\mathcal{D})
    &:=
    \omega_U(\design;\mathcal{D})
    \,
    s_{\Sigma}(\design;\mathcal{D}).
    \label{eqn:mw-acquisition}
\end{align}
This acquisition function favors designs that are simultaneously uncertain in covariance structure and promising according to the objective emulator.

When $U_0^{\mathrm{sim}}(\design;\mathcal{D})$ is not negative everywhere, one may take $\omega_U(\design;\mathcal{D})=U_0^{\mathrm{sim}}(\design;\mathcal{D})$; this is done in our subsequent numerical demonstrations. More generally, a shifted and normalized weight can be used. For example, over the candidate set $\mathcal{C}$ used to optimize the acquisition, define
\begin{align}
    \omega_U(\design;\mathcal{D})
    &:=
    \frac{
        U_0^{\mathrm{sim}}(\design;\mathcal{D})
        -
        \min_{\design' \in \mathcal{C}}
        U_0^{\mathrm{sim}}(\design';\mathcal{D})
    }{
        \max_{\design' \in \mathcal{C}}
        U_0^{\mathrm{sim}}(\design';\mathcal{D})
        -
        \min_{\design' \in \mathcal{C}}
        U_0^{\mathrm{sim}}(\design';\mathcal{D})
        +
        \epsilon
    },
    \label{eqn:objective_weight}
\end{align}
where $\epsilon>0$ is a small constant for numerical stability. This transformation preserves the ordering induced by the objective emulator over $\mathcal{C}$, ensures that the acquisition remains nonnegative, and avoids sensitivity to additive shifts of the objective.

\begin{sloppypar}
The two acquisition functions play complementary roles. The uncertainty-sampling acquisition $a_{\mathrm{us}}$ is more exploratory and is useful early in pilot sampling, when little is known about either the covariance surface or the objective surface. The mean-weighted acquisition $a_{\mathrm{mw}}$ is more exploitative and can reduce pilot-sampling effort in regions that appear unimportant for optimization. However, because $a_{\mathrm{mw}}$ depends on the objective emulator, it can be misled when $U_0^{\mathrm{sim}}(\design;\mathcal{D})$ is inaccurate. In the numerical demonstrations, we therefore use $a_{\mathrm{us}}$ during the early pilot-sampling iterations and transition to $a_{\mathrm{mw}}$ once enough high-fidelity pilot data have been collected to obtain a more reliable objective emulator.
\end{sloppypar}

\subsection{Overall algorithm}

We summarize the overall two-stage method in \cref{alg:adaptive_pilot_sampling} below. 
\begin{algorithm}[H]
\caption{
Active pilot sampling for multi-fidelity estimation in stochastic optimization.
}
\label{alg:adaptive_pilot_sampling}
\begin{algorithmic}[1]
\Require Design domain $\Design$; model ensemble $\{f_m\}_{m=0}^{M-1}$; model costs $w=[w_0,\ldots,w_{M-1}]^\top$; number of active-learning pilot-design locations $N_{\mathrm{pd}}$; number of random-input pilot samples per active-learning pilot-design location $N_{\mathrm{pilot}}$; number of outer optimization iterations $N_{\mathrm{opt}}$ for stochastic optimization; pilot-sampling acquisitions $a_{\mathrm{us}}$ and $a_{\mathrm{mw}}$.
\Ensure Approximate optimizer $\widehat{\design}^{\ast}$ and trained covariance emulator.
\vspace{2mm}

\Statex \makebox[\linewidth][c]{\underline{\textbf{Stage 1: Pilot Sampling}}}
\vspace{2mm}
\State Initialize the pilot dataset $\mathcal{D}$ using an initial space-filling set of pilot-design locations.
\State Initialize the objective surrogate, if used, with the high-fidelity pilot data.

\For{$l=1,\ldots,N_{\mathrm{pd}}$}
    \State Select the next pilot-design location
    \begin{align}
        \design_l
        \in
        \arg\max_{\design\in\Design}
        a_l(\design;\mathcal{D}),
    \end{align}
    where $a_l=a_{\mathrm{us}}$ during exploratory phase and $a_l=a_{\mathrm{mw}}$ during objective-weighted phase.

    \State Draw random-input pilot samples
    \begin{align}
        \left\{\theta_l^{(j)}\right\}_{j=1}^{N_{\mathrm{pilot}}}
        \sim
        p(\Theta).
    \end{align}

    \State Evaluate all model fidelities on the common pilot samples,
    \begin{align}
        f_m\left(\theta_l^{(j)},\design_l\right),
        \qquad
        m=0,\ldots,M-1,
        \quad
        j=1,\ldots,N_{\mathrm{pilot}}.
    \end{align}

    \State Form the sample covariance estimate $\widehat{\Sigma}(\design_l)$ using \eqref{eqn:cov}.
    \State Transform $\widehat{\Sigma}(\design_l)$ to observations $\widehat{\gamma}(\design_l)$ and $\log\widehat{\sigma}(\design_l)$ using \eqref{eqn:cov_to_corr_sigma}--\eqref{eqn:gamma_mapping}.
    \State Estimate observation-noise variances for the transformed quantities using the pilot-sample covariance sampling approximation.
    \State Update $\mathcal{D}$ with the new transformed covariance observations.
    \State Refit or update the covariance emulator conditioned on $\mathcal{D}$.
    \State Update the objective surrogate using the high-fidelity pilot data, if used.
\EndFor
\State Freeze the covariance emulator for use in the downstream stochastic optimization stage.

\algstore{myalg}
\end{algorithmic}
\end{algorithm}

\clearpage
\addtocounter{algorithm}{-1}
\begin{algorithm}[H]
\caption{Active pilot sampling for multi-fidelity estimation in stochastic optimization (Continued).
}
\begin{algorithmic}[1]
\algrestore{myalg}

\vspace{2mm}
\Statex \makebox[\linewidth][c]{\underline{\textbf{Stage 2: Stochastic Optimization}}}
\vspace{2mm}
\For{$k=1,\ldots,N_{\mathrm{opt}}$}
    \State Select the next optimization design $\design_k$ using the chosen stochastic optimization method.
    \State Query the covariance emulator at $\design_k$ to obtain a design-specific covariance estimate,
    \begin{align}
        \widehat{\Sigma}_{\mathrm{emu}}(\design_k).
    \end{align}
    \State Use $\widehat{\Sigma}_{\mathrm{emu}}(\design_k)$ and model costs $w$ to determine ACV weights and sample allocations.
    \State Evaluate the ACV estimator to obtain a noisy objective estimate
    \begin{align}
        \widehat{U}_0(\design_k).
    \end{align}
    \State Update the outer optimization method with $\widehat{U}_0(\design_k)$ and, when used, its estimated observation variance.
\EndFor

\State Return the final predicted optimizer $\widehat{\design}^{\ast}$ from the outer optimization method.
\end{algorithmic}
\end{algorithm}

The algorithm is written for an ACV estimator inside a generic outer stochastic optimization loop, but the same structure applies to other covariance-dependent multi-fidelity estimators. We use $N_{\mathrm{pd}}$ to denote the number of active-learning pilot-design locations excluding any initial space-filling pilot designs used to initialize the GP emulators, $N_{\mathrm{pilot}}$ to denote the number of random-input pilot samples drawn at each active-learning pilot-design location, and $N_{\mathrm{opt}}$ to denote the number of outer optimization iterations. At each pilot-design location, all model fidelities are evaluated on common random-input samples so that a local covariance estimate can be formed. The resulting transformed covariance observations define the pilot dataset $\mathcal{D}$ used to train the covariance emulator.

%% file: sections/4_demos.tex
\section{Numerical demonstrations}
\label{s: results}

We demonstrate the proposed pilot-sampling methods on two problems. First, we use a one-dimensional (1D) four-branch benchmark to compare covariance-emulation-based pilot sampling against single-fidelity MC, domain-averaged covariance estimation, and independent pilot sampling at each optimization iteration. Second, we apply the framework to a multi-fidelity Bayesian OED problem for sensor placement and measurement-time selection in a convection--diffusion system.

\subsection{Four-branch benchmark}\label{ss:four-branch}

\subsubsection{Problem setup}

We first demonstrate the pilot-sampling framework on a 1D benchmark stochastic optimization problem based on the four-branch limit-state function~\citep{Schueremans_Van_Gemert_2005}. We set the design domain to $\Design=[-3,3]$ and construct a three-model multi-fidelity ensemble by changing the constants $a_m$, $b_m$, $c_m$, and $d_m$ across fidelity levels. The model ensemble is defined as
\begin{align}
    f_m(\design,\Theta)
    =
    \min
    \left\{
    \begin{array}{l}
        6 + \left( \dfrac{1}{a_m} \operatorname{sgn}(\design-\Theta)|\design-\Theta|^{b_m} - \dfrac{\design+\Theta}{\sqrt{2}} \right), \\[0.3em]
        6 - \left( \dfrac{1}{a_m} \operatorname{sgn}(\design-\Theta)|\design-\Theta|^{b_m} - \dfrac{\design+\Theta}{\sqrt{2}} \right), \\[0.3em]
        \hfill 3 + \dfrac{c_m}{\sqrt{2}} + \operatorname{sgn}(\design-\Theta)|\design-\Theta|^{d_m}, \\[0.3em]
        \hfill 3 + \dfrac{c_m}{\sqrt{2}} - \operatorname{sgn}(\design-\Theta)|\design-\Theta|^{d_m}
    \end{array}
   \right\}.
    \label{eqn:four-branch}
\end{align}
The signed-power terms, such as $\operatorname{sgn}(\design-\Theta)|\design-\Theta|^{b_m}$, replace the standard squared and linear difference terms from the original four-branch function. This construction allows the exponents $b_m$ and $d_m$ to vary continuously across fidelity levels while keeping the model output real-valued. The fidelity-dependent constants and model costs are given in \cref{tab:constants}.

\begin{table}[htbp]
    \centering
    \begin{tabular}{| c | c | c c c c |}
        \hline
        \textbf{Fidelity level} $m$ & \textbf{Model cost} $w_m$ & $a_m$ & $b_m$ & $c_m$ & $d_m$ \\
        \hline\hline
        0 & 1 & 10 & 2 & 7 & 1.4 \\ \hline
        1 & $10^{-2}$ & 9 & 1.75 & 6 & 1.2 \\ \hline
        2 & $10^{-4}$ & 5 & 1.5 & 5 & 1.0 \\
        \hline
    \end{tabular}
    \caption{Fidelity-dependent parameter values and model costs for the four-branch benchmark in \eqref{eqn:four-branch}.}
    \label{tab:constants}
\end{table}

\begin{figure}[htbp]
    \centering
    \begin{subfigure}[b]{0.49\textwidth}
        \centering
        \includegraphics[width=\linewidth]{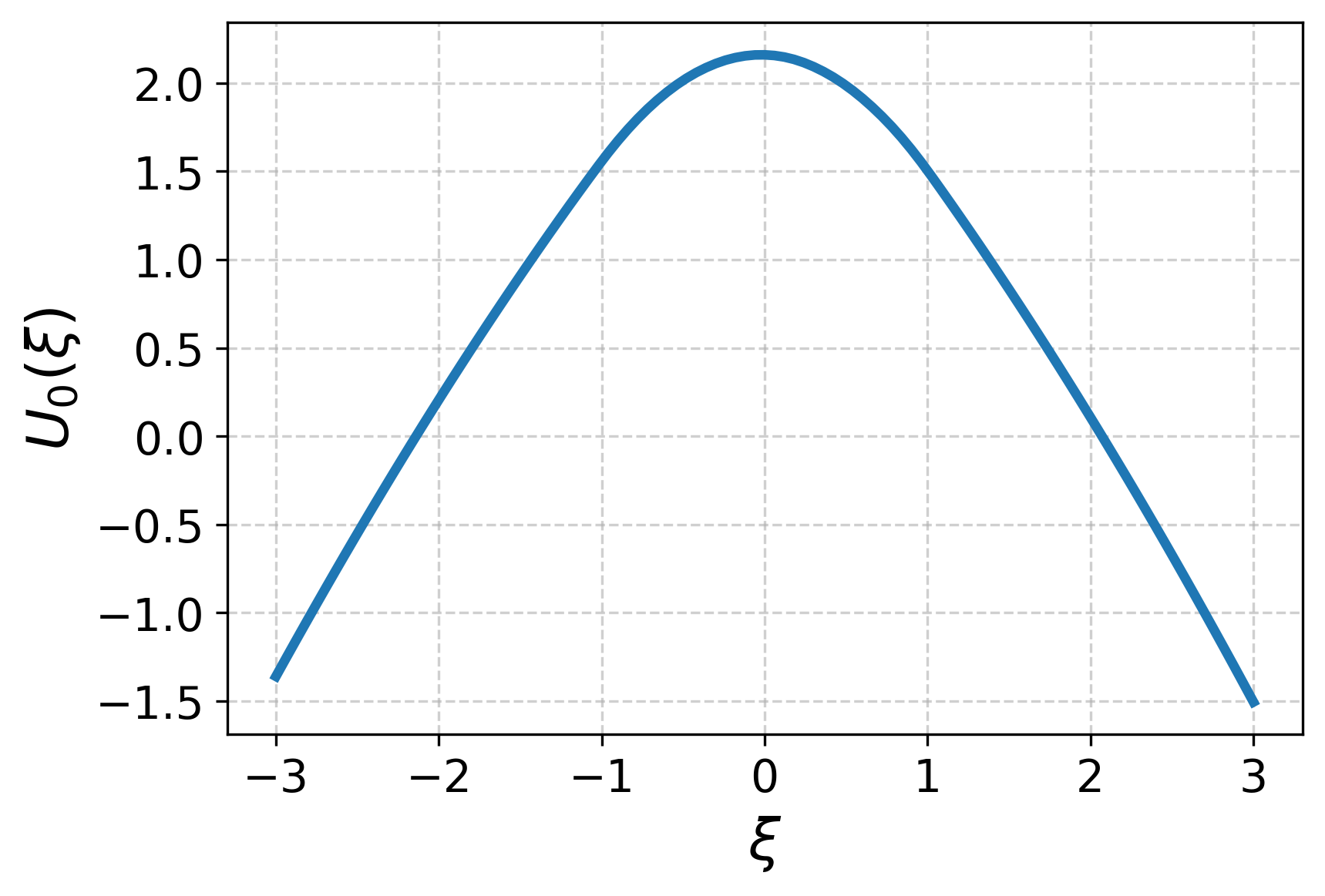}
        \caption{High-fidelity model mean $U_0(\design)$}
        \label{fig:four_branch_mean}
    \end{subfigure}
    \hfill 
    \begin{subfigure}[b]{0.47\textwidth}
        \centering
        \includegraphics[width=0.98\linewidth]{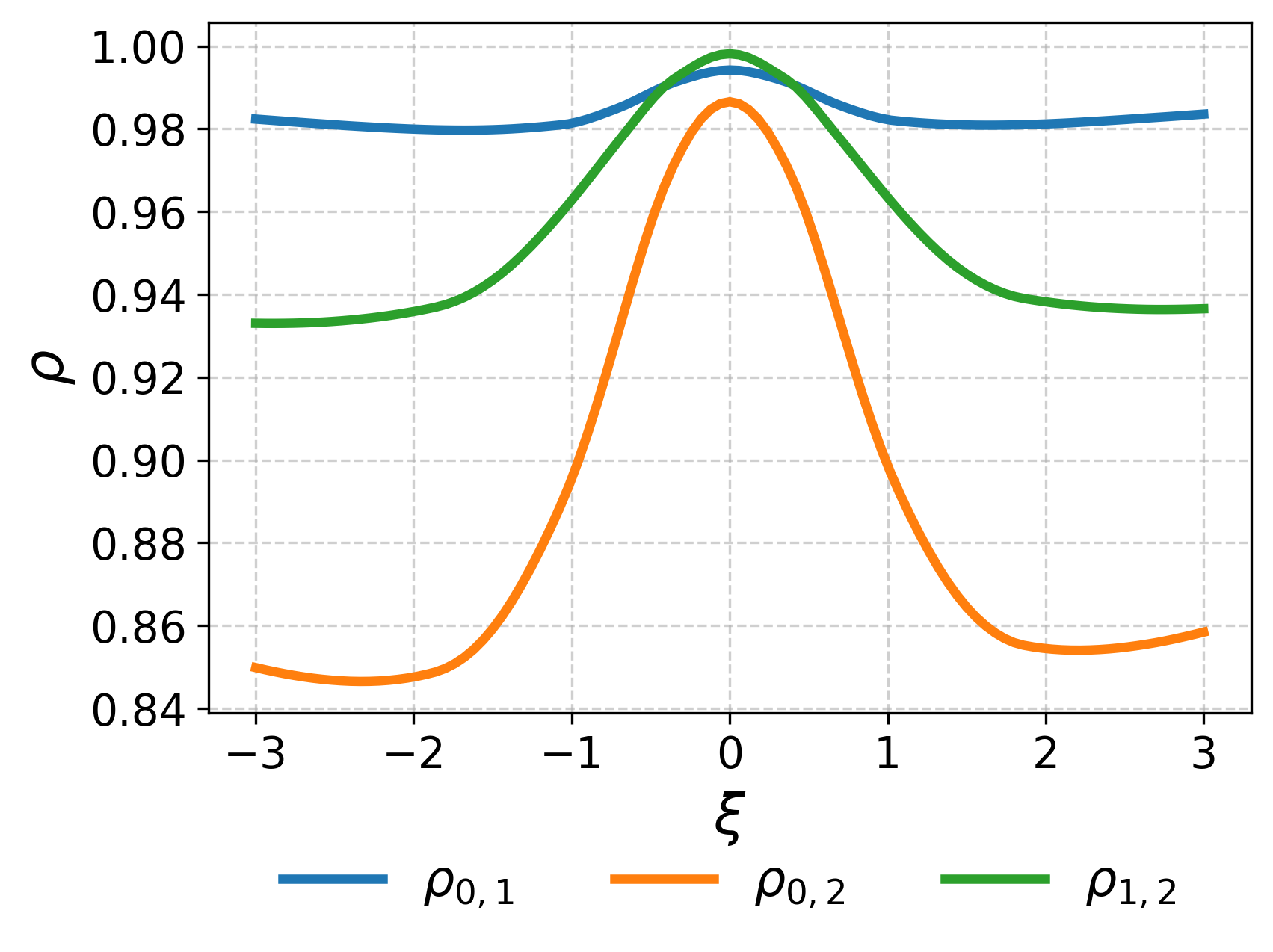}
        \caption{Multi-fidelity model correlations}
        \label{fig:four_branch_corrs}
    \end{subfigure}
    \caption{Reference quantities for the four-branch stochastic optimization benchmark. \textbf{(a)} High-fidelity mean objective $U_0(\design)$. \textbf{(b)} Pairwise model-output correlations across the design domain. The spatial variation in correlation indicates that the covariance information used to construct ACV estimators is design dependent.}
    \label{fig:four_branch_reference}
\end{figure}

We define the random input as $\Theta\sim\mathcal{U}(-6,6)$. The high-fidelity mean objective is
$U_0(\design)=\mathbb{E}_{\Theta} \left[f_0(\design,\Theta)\right]$,
and the stochastic optimization problem is 
\begin{align*}
    \design^\ast
    \in
    \arg \max_{\design\in\Design}
    U_0(\design).
\end{align*}
The reference objective curve is plotted in \cref{fig:four_branch_mean}, with the optimum, approximated using $10^6$ MC samples, at
$\left(\design^\ast,U_0(\design^\ast)\right)=(0,2.159)$.
The associated multi-fidelity model correlations are shown in \cref{fig:four_branch_corrs}. These correlations vary smoothly across the domain, indicating that the covariance information used to construct ACV estimators is design dependent. The GP regression task described in \cref{ss:gp-gamma} is then used to learn this covariance structure from independent pilot samples.

We compare four approaches for estimating $U_0$ within the stochastic optimization loop:
\begin{enumerate}
    \item \textbf{MC-SF}: a single-fidelity MC estimator \eqref{eqn:mc} using only the high-fidelity model $f_0$.

    \item \textbf{ACV-$\boldsymbol{\gamma}$-OPT}: a multi-fidelity ACV estimator \eqref{eqn:acv_formula} whose hyperparameters are optimized at each optimization query using a design-specific covariance estimate from the trained covariance emulator.

    \item \textbf{ACV-FLAT}: a multi-fidelity ACV estimator \eqref{eqn:acv_formula} whose hyperparameters are optimized using a single domain-averaged covariance matrix estimated from offline, space-filling pilot samples.

    \item \textbf{ACV-INDEP}: a multi-fidelity ACV estimator \eqref{eqn:acv_formula} whose covariance information is estimated independently at each optimization query. The number of pilot samples used at each query is set using the adaptive method from~\cite{coons_bayesian_2025}, applied at the optimum.
\end{enumerate}

Across all methods, we fix the total computational budget, including any pilot-sampling costs, and use BO with the \texttt{qLogNoisyExpectedImprovement} acquisition function~\citep{Ament2023} for the stochastic optimization stage. The methods differ only in how they obtain covariance information. \textbf{ACV-$\boldsymbol{\gamma}$-OPT} first runs an offline active-learning pilot-sampling stage to train the covariance emulator; the trained emulator is then held fixed during BO and supplies design-specific covariance estimates for ACV objective evaluations. \textbf{ACV-FLAT} uses the same offline pilot-sampling budget, but compresses the resulting pilot data into a single domain-averaged covariance estimate. \textbf{ACV-INDEP} does not use an offline pilot-sampling stage and instead estimates covariance information locally within each BO iteration. \textbf{MC-SF} performs no pilot sampling and uses only high-fidelity MC objective estimates.

This budget accounting ensures that all comparisons include the cost of obtaining covariance information. Since \textbf{MC-SF} spends no budget on pilot sampling, and \textbf{ACV-INDEP} avoids an offline pilot-sampling stage, these methods have larger per-iteration optimization budgets than \textbf{ACV-$\boldsymbol{\gamma}$-OPT} and \textbf{ACV-FLAT}.

All methods are initialized using a space-filling set of samples at two design locations, with $20$ high-fidelity model runs used to initialize the BO surrogate for $U_0$.\footnote{For \textbf{ACV-INDEP}, two additional initial design locations are queried because this method is more susceptible to under-exploration.} 
For \textbf{ACV-$\boldsymbol{\gamma}$-OPT}, these initial samples are also used to initialize the $\gamma$-GPs. To implement active-learning pilot sampling for \textbf{ACV-$\boldsymbol{\gamma}$-OPT}, we use the uncertainty-sampling acquisition function $a_{\mathrm{us}}$ in \eqref{eqn:us-acquisition} for the first half of the pilot-sampling iterations, then switch to the mean-weighted acquisition function $a_{\mathrm{mw}}$ in \eqref{eqn:mw-acquisition} for the second half. During pilot sampling, the high-fidelity samples are also used to form noisy single-fidelity MC estimates of $U_0$, which update the BO surrogate.

\subsubsection{Results}\label{sss:results}

\paragraph{An illustrative optimization run}

We first present a single optimization run to illustrate the active-learning pilot-sampling procedure, focusing on \textbf{ACV-$\boldsymbol{\gamma}$-OPT}. For this run, we set $N_{\mathrm{pd}}=12$ and $N_{\mathrm{pilot}}=32$, giving a pilot cost of $32\sum_{m=0}^{M-1} w_m = 32.3232$ at each pilot-design location. The total computational budget is fixed to $B=2988.2824$ for each method. After accounting for a GP initialization using $2$ pilot designs with $20$ samples each, this budget allows $N_{\mathrm{opt}}=40$ mean-optimization iterations, with per-iteration budgets of $74.7$ for \textbf{MC-SF}, $64$ for \textbf{ACV-$\boldsymbol{\gamma}$-OPT} and \textbf{ACV-FLAT}, and $73.7$ for \textbf{ACV-INDEP}. These budgets differ because \textbf{ACV-$\boldsymbol{\gamma}$-OPT} and \textbf{ACV-FLAT} allocate part of the total budget to offline pilot sampling, while \textbf{ACV-INDEP} allocates budget to local pilot sampling within each optimization iteration.

\begin{figure}[htbp]
    \centering
    \includegraphics[width=\linewidth]{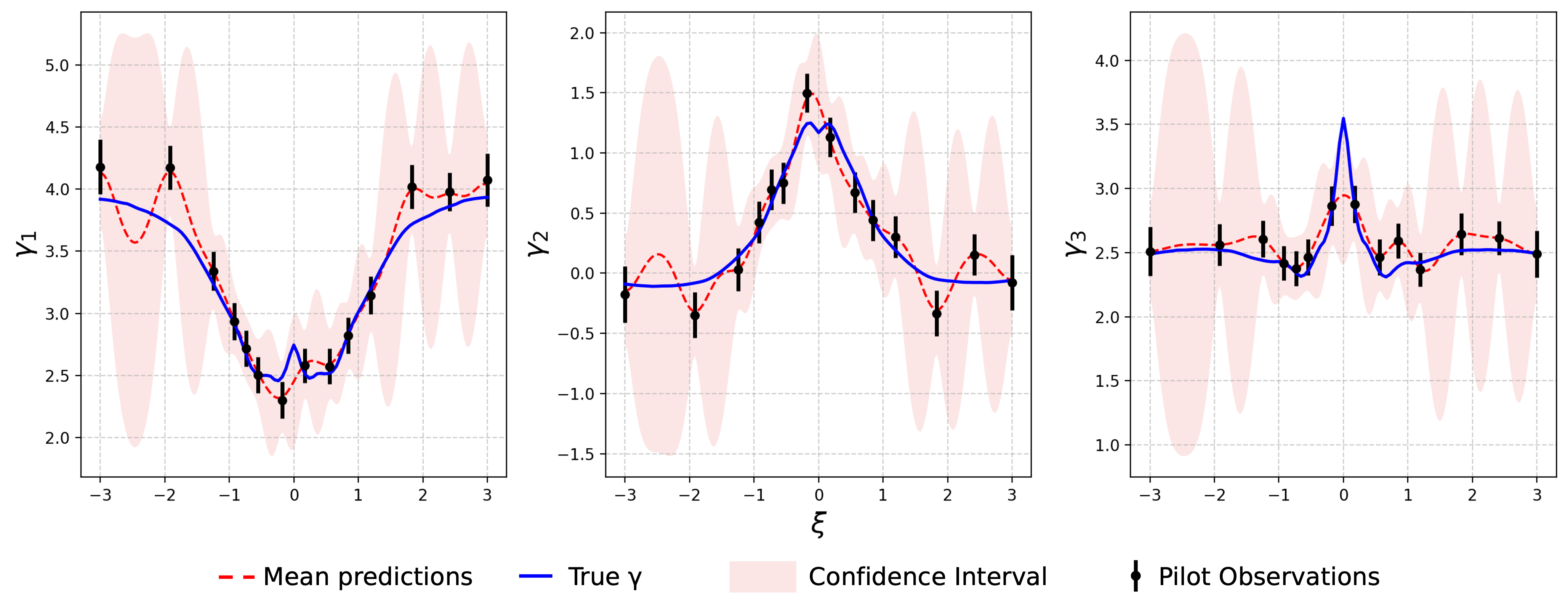}
    
    \caption{Posterior $\gamma$-GPs after active-learning pilot sampling for an illustrative \textbf{ACV-$\boldsymbol{\gamma}$-OPT} run on the four-branch benchmark. }
    \label{fig:gammas_post}
\end{figure}

The \textbf{ACV-$\boldsymbol{\gamma}$-OPT} method begins with active-learning pilot sampling, targeting both posterior uncertainty in the $\gamma$-GPs and promising regions of the $U_0$-GP. This stage produces an initialized $U_0$-GP and a resolved set of $\gamma$-GPs, shown in \cref{fig:gammas_post}. The posterior mean of the transformed covariance emulator is then mapped back to model correlations and used to set the ACV hyperparameters during the subsequent mean-optimization loop.

We next compare the mean-optimization behavior of \textbf{ACV-$\boldsymbol{\gamma}$-OPT} and \textbf{MC-SF}. \Cref{fig:four_branch_demo} shows the posterior $U_0$-GPs and their predicted maximizers $\widehat{\design}^{\ast}$ at $20\%$ ($8$ iterations), $50\%$ ($20$ iterations), and $100\%$ ($40$ iterations) of the mean-optimization loop.\footnote{These checkpoints differ from the budget checkpoints in \cref{tab:reduction_factors}, which include pilot-sampling costs when tracking the total budget used. For example, after $8$ mean-optimization iterations, \textbf{ACV-$\boldsymbol{\gamma}$-OPT} has used more total budget than \textbf{MC-SF} because \textbf{MC-SF} does not require pilot sampling.} The posterior uncertainty of the $U_0$-GP is substantially smaller when using ACV because each objective observation has lower estimator variance. This lower noise leads to more accurate predicted maximizers, especially later in the optimization process, where \textbf{MC-SF} struggles to converge reliably. Although the improvement is visually subtle after the eighth iteration, the predicted maximizer for \textbf{ACV-$\boldsymbol{\gamma}$-OPT} continues to move toward the true maximizer at $\design^\ast=0$.

In this illustrative run, the offline pilot-sampling stage accounts for just under $13\%$ of the total optimization budget, yet enables clear improvements over single-fidelity MC and independent pilot sampling. The high-fidelity pilot samples are used not only for covariance estimation but also to inform the $U_0$-GP, albeit noisily.

\paragraph{Summary of full optimization results}

To compare the methods across optimization settings, we repeat $30$ trials of each method over $12$ hyperparameter configurations. Across all configurations, the total computational budget and the number of mean-optimization iterations are fixed. We vary the per-iteration optimization budget for \textbf{ACV-$\boldsymbol{\gamma}$-OPT} and \textbf{ACV-FLAT} across
$B_{\mathrm{iter}} \in \{32,64,128\}$,
the number of pilot-design locations across
$N_{\mathrm{pd}} \in \{6,12\}$,
and the number of random-input pilot samples per pilot-design location across
$N_{\mathrm{pilot}} \in \{16,32\}$. 
\Cref{tab:reduction_factors} reports regret reduction factors relative to the single-fidelity baseline \textbf{MC-SF}, averaged over all trials and over all hyperparameter configurations not shown in the table. Full optimization results are provided in Appendix~\ref{app:four_branch_results}.

\begin{table}[ht]
    \centering
    \small
    \setlength{\tabcolsep}{4pt}
    \caption{Four-branch optimization results: regret reduction factor relative to the single-fidelity baseline \textbf{MC-SF} for $N_{\mathrm{pd}}=6$ and $N_{\mathrm{pd}}=12$ pilot-design locations, averaged over all other optimization hyperparameter configurations/setups.}
    \label{tab:reduction_factors}
    \resizebox{\linewidth}{!}{\begin{tabular}{|c|l|rrr|}
        \hline
        \textbf{Setup} & \textbf{Method} & \textbf{20\% Budget} & \textbf{50\% Budget} & \textbf{100\% Budget} \\
        \hline\hline
        \multirow{3}{*}{$N_{\mathrm{pd}}=6$} 
        & ACV-$\boldsymbol{\gamma}$-OPT & $1.8\times$ & $\mathbf{7.9\times}$ & $\mathbf{25.7\times}$ \\
        & ACV-FLAT & $\mathbf{2.1\times}$ & $5.5\times$ & $8.0\times$ \\
        & ACV-INDEP & $0.6\times$ & $3.3\times$ & $8.5\times$ \\
        \hline
        \multirow{3}{*}{$N_{\mathrm{pd}}=12$} 
        & ACV-$\boldsymbol{\gamma}$-OPT & $\mathbf{2.6\times}$ & $\mathbf{13.6\times}$ & $\mathbf{35.8\times}$ \\
        & ACV-FLAT & $2.2\times$ & $8.3\times$ & $16.4\times$ \\
        & ACV-INDEP & $0.7\times$ & $3.4\times$ & $10.0\times$ \\
        \hline
    \end{tabular}}
\end{table}

All multi-fidelity approaches outperform \textbf{MC-SF} at termination with $8$--$35\times$ reductions in optimization regret. The best performance comes from methods that share covariance information across the design domain: \textbf{ACV-FLAT} through a single domain-averaged covariance estimate, and \textbf{ACV-$\boldsymbol{\gamma}$-OPT} through a design-dependent covariance emulator. One of these two methods performs best across every hyperparameter configuration and budget checkpoint tested.

\sloppy The relative performance of the ACV methods changes over the optimization process. \textbf{ACV-FLAT} is competitive early, and sometimes outperforms \textbf{ACV-$\boldsymbol{\gamma}$-OPT}, because averaging covariance information across the domain reduces covariance-estimation noise. As the optimization loop converges, however, local covariance information becomes more important, and \textbf{ACV-$\boldsymbol{\gamma}$-OPT} provides the largest regret reductions. In contrast, \textbf{ACV-INDEP} performs poorly early because it repeatedly estimates covariance information from limited local pilot data, but becomes more competitive near convergence as local covariance accuracy becomes increasingly valuable. These trends indicate that \textbf{ACV-$\boldsymbol{\gamma}$-OPT} provides the desired balance: it shares covariance information across the design domain while still adapting ACV hyperparameters to local covariance structure.

The results also show that investing more budget in pilot sampling can improve long-term optimization performance. Using $N_{\mathrm{pd}}=12$ yields better regret reduction than $N_{\mathrm{pd}}=6$ for the ACV-based methods across all budget checkpoints on average. This reflects the value of learning more accurate covariance information, which improves both the ACV estimates and the observation-noise estimates supplied to the $U_0$-GP. For simpler covariance surfaces or smaller total budgets, the optimal pilot-sampling investment may be lower. Determining the best division of budget between covariance learning and mean-objective optimization remains problem- and optimizer-dependent, and is left for future work.

\subsection{Application to Bayesian optimal experimental design}\label{ss:OED}

\subsubsection{Problem setup}

We next demonstrate the proposed framework on a Bayesian OED problem involving a two-dimensional (2D) convection--diffusion model. Related contaminant-source inversion problems were studied in the context of sequential OED in~\cite{shen_bayesian_2023,shen_variational_2025}; here, we adapt the setup by modifying the diffusion strength, velocity field, and domain bounds. The physical system consists of two contaminant sources at unknown locations. These sources release contaminant into a spatially varying velocity field while the contaminant diffuses over time.

Let $G(x,t; \Theta)$ denote the contaminant concentration at spatial location $x=[x_1,x_2]^{\top}\in[-1,1]^2$ and time $t$. The high-fidelity dynamics are
\begin{align}
    \frac{\partial G(x,t;\Theta)}{\partial t}
    +
    v(x;\Theta)\cdot \nabla G(x,t;\Theta)
    &=
    k\nabla^2 G(x,t;\Theta)
    +
    S(x;\Theta),
    \label{eq:conv_diff_2d_temporal}
    \\
    x\in[-1,1]^2,\quad t>0, \nonumber
\end{align}
with initial condition $G(x,0; \Theta)=0$ and homogeneous Neumann boundary conditions on all sides of the square domain. Here, $k$ is the diffusion coefficient, $S$ is the source term, and $v=[v_1,v_2]^\top$ is a time-independent convective velocity field.

The two-source term is modeled as
\begin{align}
    S(x;\Theta)
    =
    \sum_{i=1}^{2}
    \frac{s}{2\pi h^2}
    \exp\left(
        -
        \frac{
            (\Theta_{i,x}-x_1)^2
            +
            (\Theta_{i,y}-x_2)^2
        }{
            2h^2
        }
    \right),
    \label{eq:source_fn}
\end{align}
where $s$ and $h$ are the source strength and width, respectively. The velocity field is inspired by~\cite{fries_lasdi:_2022} and is defined as
\begin{align}
    v(x;\Theta)
    =
    \begin{bmatrix}
        v_1(x;\Theta) \\
        v_2(x;\Theta)
    \end{bmatrix}
    =
    \begin{bmatrix}
        \dfrac{\Theta_{\beta}}{2}\pi(1-x_1^2)x_2 - \zeta x_1 \\
        -\dfrac{\Theta_{\beta}}{2}\pi(1-x_2^2)x_1 - \zeta x_2
    \end{bmatrix}.
    \label{eq:vel_field}
\end{align}
The uncertain parameters are
\begin{align}
    \Theta
    =
    \left[\Theta_{1,x},\Theta_{1,y},\Theta_{2,x},\Theta_{2,y},\Theta_{\beta}\right]^{\top}.
\end{align}
We use independent priors
\begin{align}
    \Theta_{1,x}    &\sim
    \mathcal{U}(-1,0),
    &
    \Theta_{2,x}
    &\sim
    \mathcal{U}(0,1),
    &
    \Theta_{1,y},\Theta_{2,y} &\sim \mathcal{U}(-1, 1),
    &
    \Theta_{\beta}
    &\sim
    \mathcal{U}(5,12).
\end{align}
The remaining physical parameters are fixed at $s=2$, $h=0.05$, $\zeta=2.5$, and $k=0.25$.
The role of design variable $\design$ (e.g., sensor location $x$ or sensing time $t$) will depend on the particular case setup introduced later.

For a given experimental design $\design\in\Design$, let $Y$ denote the corresponding noisy observation vector. We define the pointwise information utility as the log likelihood-to-evidence ratio,
\begin{align}
    u(Y,\Theta;\design)
    :=
    \log
    \left(
        \frac{
            p(Y|\Theta,\design)
        }{
            p(Y|\design)
        }
    \right).
    \label{eqn:eig_utility}
\end{align}
The Bayesian OED problem is to choose a design that maximizes the expected information gain (EIG),
\begin{align}
    U_{\mathrm{EIG}}(\design)
    &=
    \mathbb{E}_{Y,\Theta|\design}
    \left[
        u(Y,\Theta;\design)
    \right].
    \label{eqn:EIG}
\end{align}
Equivalently, the EIG is the mutual information between $\Theta$ and $Y$ under design $\design$, or the expected reduction in uncertainty about $\Theta$ after observing data from that experiment.

To allow cost-aware design, we define
\begin{align}
    U_c(\design)
    :=
    \frac{
        U_{\mathrm{EIG}}(\design)
    }{
        C(\design)
    },
    \label{eqn:cost-EIG}
\end{align}
where $C(\design)$ is a known experimental cost function. If experiments have equal cost across designs, then $C(\design)=1$ and \eqref{eqn:cost-EIG} reduces to the standard EIG objective. The cost-aware OED problem is
\begin{align}
    \design^\ast
    \in
    \arg\max_{\design\in\Design}
    U_c(\design).
    \label{eqn:oed_opt}
\end{align}

For nonlinear models, $U_{\mathrm{EIG}}(\design)$ is generally unavailable in closed form. A standard estimator is the nested MC (NMC) estimator
\begin{align}
    \widehat{U}_{\mathrm{NMC}}(\design)
    :=
    \frac{1}{N_{\mathrm{out}}}
    \sum_{i=1}^{N_{\mathrm{out}}}
    \left\{
        \log p(y^{(i)}|\theta^{(i)},\design)
        -
        \log
        \left[
            \frac{1}{N_{\mathrm{in}}}
            \sum_{j=1}^{N_{\mathrm{in}}}
            p(y^{(i)}|\theta^{(i,j)},\design)
        \right]
    \right\}.
    \label{eqn:NMC}
\end{align}
Here, $\theta^{(i)}$ are i.i.d. samples from the prior $p(\theta)$, $y^{(i)}$ are corresponding samples from the likelihood $p(y|\theta^{(i)},\design)$, and $\theta^{(i,j)}$ are additional prior samples used to estimate the evidence term. The inner and outer sample sizes are $N_{\mathrm{in}}$ and $N_{\mathrm{out}}$, respectively. At a fixed design, the NMC estimator requires $\mathcal{O}(N_{\mathrm{out}}N_{\mathrm{in}})$ likelihood evaluations and, therefore, $\mathcal{O}(N_{\mathrm{out}}N_{\mathrm{in}})$ high-fidelity forward model evaluations.

To reduce this cost, we use the ACV-based MF-EIG estimator introduced in~\cite{coons_multi-fidelity_2025}. This estimator applies ACV to a set of utility models $u_m$, each associated with a forward model $G_m$. The high-fidelity utility model $u_0$ is constructed from the high-fidelity forward model $G_0$, while the low-fidelity utility models $u_1$ and $u_2$ are constructed from low-fidelity forward models $G_1$ and $G_2$, to be introduced in \cref{ss:oed-models}. Each utility model approximates $u$ in \eqref{eqn:eig_utility}, but also absorbing the inner loop in \eqref{eqn:NMC} as part of the utility model. ACV is then applied across the outer-loop utility evaluations, reducing variance while remaining unbiased with respect to the high-fidelity utility model. In the numerical experiments, we compare the single-fidelity NMC baseline \textbf{NMC-SF} against MF-EIG estimators using a domain-averaged covariance matrix, \textbf{MFEIG-FLAT}, and a design-dependent covariance emulator, \textbf{MFEIG-OPT}.

We use two OED cases to highlight different aspects of the proposed framework. Case 1 is an all-at-once spatial sensor-placement problem, which demonstrates the value of local covariance information for MF-EIG estimation even when an iterative optimization loop is not needed. Case 2 is an iterative measurement-time design problem, which directly exercises the proposed adaptive pilot-sampling and optimization workflow. The two cases are summarized in \cref{tab:contaminant_source_inversion}.

\paragraph{Case 1: all-at-once estimation for sensor placement} A mobile sensor measures contaminant concentration at a spatial location $x\in[-1,1]^2$ at a fixed time $t'=0.325$. The goal is to choose the sensor location $x$ that maximizes information about the source parameters $\Theta$. We assume equal experimental costs, so $C(\design)=1$.

For a sensor placed at location $x$, the noisy observation is
\begin{align}
    Y
    =
    G(x,t'; \Theta)
    +
    \mathcal{E},
    \label{eqn:obs_loc}
\end{align}
where $\mathcal{E}\sim\mathcal{N}(0,\sigma_{\mathcal{E}}^2)$. In this case, the design variable is the sensor location, so $\design=x$ and $\Design=[-1,1]^2$.

This case is what we call an ``all-at-once'' estimation problem because here, a single forward simulation produces the contaminant field over the full spatial domain. After sampling the observation noise, the resulting field can be used to evaluate candidate observations at all sensor locations. Thus, EIG estimates across the spatial domain can be formed without independently querying each design in an outer optimization loop. This does not require iterative pilot sampling and prevents the ACV sample allocation from being changed separately at each design. Nevertheless, the case is useful for demonstrating the OED setup, the variance reduction of MF-EIG estimation, and the benefit of tailoring MF-EIG \emph{weights} to local covariance information even without altering the sample allocation.

\paragraph{Case 2: iterative estimation for sensor measurement time}
We next fix the sensor location at $x'=[-0.8,-0.2]^{\top}$ and choose the measurement time $t_k\in[0.1,0.6]$. In this case, later measurements incur a linear cost penalty, $C(\design) = 0.1+\design$, where $\design=t_k$. The noisy observation is
 \begin{align}
    Y_k
    =
    G(x',t_k; \Theta)
    +
    \mathcal{E}_k,
    \label{eqn:obs_time}
\end{align}
where $\mathcal{E}_k\sim\mathcal{N}(0,\sigma_{\mathcal{E}}^2)$. Unlike Case 1, each queried design corresponds to a single measurement time. We therefore use this case to demonstrate the full iterative workflow, including active-learning pilot sampling, covariance emulation, and outer stochastic optimization.

For simplicity, we assume that each model returns only the contaminant concentration at the requested measurement time and that the computational cost of each model is independent of $t_k$.\footnote{In practice, because the lower-fidelity surrogates generate state predictions at a fixed cadence up to $t_k$, additional cost savings may be possible by reusing intermediate predictions over the optimization loop.}

\subsubsection{Model ensemble}\label{ss:oed-models}

To enable MF-EIG estimation, we construct a model ensemble consisting of one high-fidelity model, $G_0$, and two lower-fidelity approximations, $G_1$ and $G_2$. The associated utility models are denoted by $u_0$, $u_1$, and $u_2$, and ACV is applied directly to these utility models as in~\cite{coons_multi-fidelity_2025}.

The intended high-fidelity model for this problem is the finite-difference solver for \eqref{eq:conv_diff_2d_temporal}. Using the settings in \cref{tab:contaminant_source_inversion}, this solver uses a uniform grid with $\Delta x_1=\Delta x_2=0.02$ and a second-order fractional-step time-marching scheme with step size $\Delta t=5.0\times 10^{-4}$ up to final time $t_M=0.6$. In a practical OED study, this finite-difference solver would be used directly as the high-fidelity model. However, for benchmarking purposes, we also need sufficiently accurate reference EIG and optimization results against which to compare the proposed estimators. Generating these references with the finite-difference solver would require a prohibitively large number of high-fidelity model evaluations.

To make such comparisons feasible, we train a multilayer perceptron (MLP) proxy using a large dataset generated from the finite-difference solver. We then use this proxy as the high-fidelity model $G_0$ in the numerical experiments. This choice makes it possible to run many optimization trials and to compute reference solutions, while preserving the input-output behavior of the finite-difference model to high accuracy. Importantly, when assigning model costs, we use the cost of the finite-difference solver rather than the runtime of the MLP proxy. Thus, the computational budget reflects the intended high-fidelity model that would be used in practice, while the proxy is used only to make benchmarking and repeated evaluation tractable. Further details on the high-fidelity proxy are provided in \cref{app:proxy}.

\begin{table}[htbp]
    \centering
    \small
    \setlength{\tabcolsep}{4pt}
    \resizebox{\linewidth}{!}{\begin{tabular}{|l|c|c|}
        \hline
         & \textbf{Case 1} & \textbf{Case 2} \\
        \hline\hline
        Source priors
        &
        \multicolumn{2}{c|}{
            $\Theta_{1,x}\sim\mathcal{U}(-1,0)$, \ 
            $\Theta_{2,x}\sim\mathcal{U}(0,1)$, \ 
            $\Theta_{1,y},\Theta_{2,y} \sim \mathcal{U}(-1, 1)$
        } \\
        \hline
        Velocity prior
        &
        \multicolumn{2}{c|}{
            $\Theta_{\beta}\sim\mathcal{U}(5,12)$
        } \\
        \hline
        Noise scale
        &
        \multicolumn{2}{c|}{$\sigma_{\mathcal{E}}=0.1$} \\
        \hline
        Design variable
        &
        Sensor location $\design=x\in[-1,1]^2$
        &
        Measurement time $\design=t_k\in[0.1,0.6]$ \\
        \hline
        Fixed setting
        &
        Measurement time $t'=0.325$
        &
        Sensor location $x'=(-0.8,-0.2)$ \\
        \hline
        Cost function
        &
        $C(\design)=1$
        &
        $C(\design)=0.1+\design$ \\
        \hline
    \end{tabular}}
    \caption{Setup of the two OED cases for contaminant measurement in a convection--diffusion field. Case 1 optimizes the spatial sensor location at a fixed measurement time, while Case 2 optimizes the measurement time at a fixed sensor location.}
    \label{tab:contaminant_source_inversion}
\end{table}

The two low-fidelity models, $G_1$ and $G_2$, are Fourier neural operator (FNO) surrogates~\citep{Li2020}. We generate $250$ full trajectories from the finite-difference solver and save snapshots up to $t_M=0.6$. Each FNO takes as input the current concentration field, the source term, the velocity components, and spatial coordinates, and predicts the concentration field at the next time step. More specifically, the autoregressive mapping is from
\begin{align}
    \left\{
        G(x,t_{k-1};\Theta),
        S(x;\Theta),
        v_1(x;\Theta),
        v_2(x;\Theta),
        x
    \right\}
    \quad
    \text{to}
    \quad
    G(x,t_k;\Theta).
\end{align}
When a prediction is needed at a time not included in the saved snapshots, linear interpolation is used between the nearest rollout times.

The FNO architecture first lifts the multichannel input into a higher-dimensional representation using a shallow fully connected layer. It then applies a sequence of spectral convolution layers and nonlinear activations before projecting back to the output space. The spectral convolutions provide an efficient parameterization in Fourier space on the uniform grid, reducing the cost of forward prediction relative to the high-fidelity solver.

We vary the network architecture and temporal resolution of the training snapshots to construct the model hierarchy:
\begin{enumerate}
    \item The medium-fidelity FNO, denoted FNO-M or $G_1$, uses four alternating spectral and regular convolutional layers. It is trained for $40$ epochs using trajectory snapshots spaced by $\Delta t_{\mathrm{train}}=0.05$.

    \item The low-fidelity FNO, denoted FNO-L or $G_2$, uses two spectral and convolutional layers. It is trained for $20$ epochs using trajectory snapshots spaced by $\Delta t_{\mathrm{train}}=0.1$.
\end{enumerate}

After training, all models are treated as black-box forward models for the OED problem. For a new parameter sample $\theta^\ast$, predictions are generated by first assembling the corresponding source term $S(x;\theta^\ast)$ and velocity field $v(x;\theta^\ast)$, then using them to evaluate the selected model. Model sizes and single-step run times are summarized in \cref{tab:surrogate_summary}. Example predictions at $t_k=0.325$ for two realizations of $\Theta$ are shown in \cref{fig:mlp_fno_fnolf_comparison} (Appendix~\ref{app:ad-vec-model-realizations}).

\begin{table}[htbp]
    \centering
    \begin{tabular}{|c|c|c|}
        \hline
        \textbf{Model} & \textbf{Parameters} & \textbf{Run time [s]} \\
        \hline\hline
        High-fidelity MLP ($G_0$) & 18,871 & 31.48 \\
        \hline
        FNO-M ($G_1$) & 465,437 & 2.039 \\
        \hline
        FNO-L ($G_2$) & 234,197 & 0.65 \\
        \hline
    \end{tabular}
    \caption{Model sizes and run times for the high-fidelity proxy model and the two low-fidelity FNO surrogates.}
    \label{tab:surrogate_summary}
\end{table}

\subsubsection{Results}
\label{sss:oed-results}

We evaluate each method by its performance on the OED optimization task. For a predicted optimizer $\widehat{\design}^{\ast}$, we define the cost-aware EIG regret
\begin{align}
    R(\widehat{\design}^{\ast})
    :=
    \left|
        U_c(\widehat{\design}^{\ast})
        -
        U_c(\design^\ast)
    \right|,
    \label{eqn:regret}
\end{align}
where $\design^\ast$ is the reference optimizer. This regret measures the suboptimality of the design selected by a particular estimation or optimization procedure, in the same units as the cost-aware objective.

To compute reference regrets, we construct an oracle EIG surface using the high-fidelity model in an NMC estimator with $10^5$ inner-loop and outer-loop samples for each EIG evaluation. \Cref{fig:reference_util} (Appendix~\ref{app:advec-ref-case-1}) shows the resulting reference EIG surface for Case 1, with the reference optimum marked at $\design^\ast=[-0.8,-0.2]^\top$.

\paragraph{Case 1: sensor placement}

In Case 1, the design is the spatial sensor location at a fixed measurement time. Because a single forward solve produces the full contaminant field, EIG estimates can be formed across the spatial design domain without an iterative outer optimization loop. We therefore use this case to test whether design-dependent covariance information improves MF-EIG estimation even when the ACV sample allocation is fixed across the domain.

We compare three estimators. The first is the single-fidelity nested MC baseline, \textbf{NMC-SF}. The second is \textbf{MFEIG-FLAT}, which uses a domain-averaged covariance matrix to determine both the ACV sample allocation and ACV weights. The third is \textbf{MFEIG-ADAPT}, which uses the same domain-averaged covariance matrix to determine the ACV sample allocation, but adapts the ACV weights across the design domain using local covariance estimates. Both MF-EIG estimators use $1000$ inner-loop samples for all utility models and share inner-loop samples across utility-model fidelities to increase inter-model correlations, following~\cite{coons_multi-fidelity_2025}.

The computational budget is set so that \textbf{NMC-SF} can use $25$ outer-loop MC samples, each with $1000$ fresh inner-loop samples. The MF-EIG estimators are given the same total budget. This setting reflects a restrictive but plausible budget in which roughly $25{,}000$ high-fidelity model evaluations are affordable. 

\begin{figure}[htbp]
    \centering
    \includegraphics[width=1\linewidth]{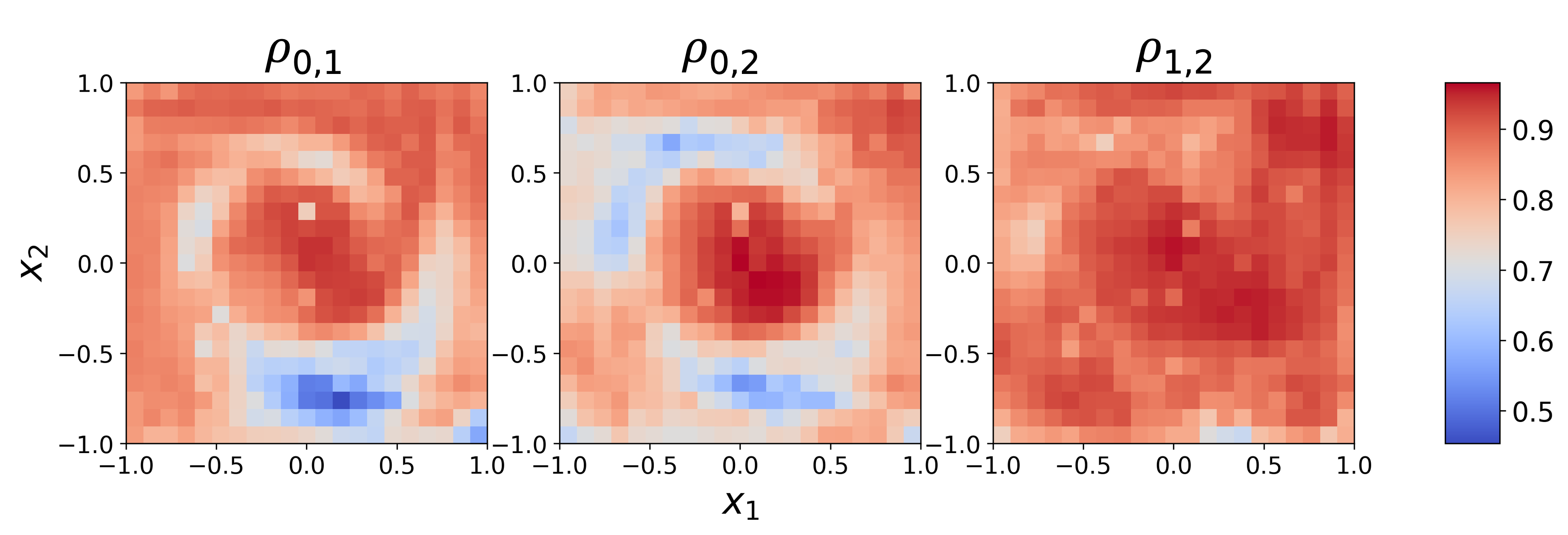}
    \caption{Utility-model correlations used to construct ACV hyperparameters for the design-adapted \textbf{MFEIG-ADAPT} estimator and the domain-averaged \textbf{MFEIG-FLAT} estimator in Case 1.}
    \label{fig:model_util_corrs}
\end{figure}

Before evaluating the MF-EIG estimators, we perform an offline pilot-sampling stage using $300$ outer-loop pilot samples to estimate utility-model covariances and determine ACV hyperparameters. The resulting spatial utility correlations at $t'=0.325$ are shown in \cref{fig:model_util_corrs}. The two low-fidelity utility models are strongly correlated with one another across the domain, while the correlations between the high-fidelity utility model and each low-fidelity utility model vary more substantially. Both \textbf{MFEIG-FLAT} and \textbf{MFEIG-ADAPT} exploit these correlations for variance reduction, but only \textbf{MFEIG-ADAPT} adapts the ACV weights to the spatial variation in covariance structure.

We report average regrets, average estimator variances, and estimator variances at the reference optimum in \cref{tab:case1}. Results are averaged over $60$ independent trials of a grid search over the spatial design grid. Both MF-EIG estimators reduce regret relative to \textbf{NMC-SF}, because the reduced estimator variance makes it easier for the grid search to distinguish among high-performing sensor locations. The difference between \textbf{MFEIG-ADAPT} and \textbf{MFEIG-FLAT} is smaller than in a fully iterative optimization setting, because the grid-search objective estimates are formed across the full domain and the estimator noise is partially smoothed spatially. Nevertheless, \textbf{MFEIG-ADAPT} achieves the lowest regret and the lowest estimator variances. This suggests that local covariance information is useful even when only the ACV weights, rather than both weights and sample allocations, are adapted across the design domain.

\begin{table}[htbp]
    \centering
    \small
    \setlength{\tabcolsep}{4pt}
    \caption{Estimator and optimization results for Case 1, the sensor-placement problem. Regrets are computed using \eqref{eqn:regret}; all values are averaged over $60$ independent grid-search trials.}
    \label{tab:case1}
    \resizebox{\linewidth}{!}{\begin{tabular}{|c|c|c|c|}
        \hline
        \textbf{Estimator}
        &
        \textbf{Average regret}
        &
        \begin{tabular}[c]{@{}c@{}}
            \textbf{Average} \\
            \textbf{estimator variance}
        \end{tabular}
        &
        \begin{tabular}[c]{@{}c@{}}
            \textbf{Estimator variance} \\
            \textbf{at true optimum}
        \end{tabular}
        \\
        \hline\hline
        NMC-SF
        &
        $7.51 \times 10^{-2}$
        &
        $4.73 \times 10^{-2}$
        &
        $4.69 \times 10^{-2}$
        \\
        \hline
        MFEIG-ADAPT
        &
        $\mathbf{4.55 \times 10^{-2}}$
        &
        $\mathbf{2.27 \times 10^{-2}}$
        &
        $\mathbf{2.40 \times 10^{-2}}$
        \\
        \hline
        MFEIG-FLAT
        &
        $5.35 \times 10^{-2}$
        &
        $2.37 \times 10^{-2}$
        &
        $2.44 \times 10^{-2}$
        \\
        \hline
    \end{tabular}}
\end{table}

\paragraph{Case 2: sensor measurement time}

In Case 2, the design is the measurement time at a fixed sensor location. Different from Case 1, each queried design requires a separate EIG estimate, making this problem a direct test of the full iterative workflow to adapt ACV hyperparameters across the design domain. We compare the proposed covariance-emulation approach, \textbf{MFEIG-$\boldsymbol{\gamma}$-OPT}, against the single-fidelity baseline \textbf{NMC-SF}. For \textbf{MFEIG-$\boldsymbol{\gamma}$-OPT}, the covariance emulator is queried at each optimization design to determine ACV hyperparameters. Both \textbf{NMC-SF} and \textbf{MFEIG-$\boldsymbol{\gamma}$-OPT} use $1000$ inner-loop samples for the high-fidelity utility model, and smaller inner-loop sample sizes of $500$ and $100$ for the low-fidelity utility models $u_1$ and $u_2$, respectively.

This case is substantially more expensive than Case 1 because each design query requires fresh EIG estimation and therefore fresh forward model evaluations. Due to this cost, we repeat \textbf{MFEIG-$\boldsymbol{\gamma}$-OPT} for $10$ independent trials. The proxy high-fidelity model allows us to run the \textbf{NMC-SF} baseline for $50$ independent trials.

\Cref{fig:case02_pilot_reference} (Appendix~\ref{app:advec-ref-case-2}) shows reference quantities for the measurement-time design problem. The cost-aware EIG curves in \cref{fig:case02_pilot_utils} show that the low-fidelity utility models do not reproduce the high-fidelity objective surface exactly. In particular, the FNO-M utility has a noticeably biased optimum, while the FNO-L utility more closely matches the high-fidelity maximizer but still differs in shape, especially for later measurement times. Directly optimizing either low-fidelity EIG surface could therefore lead to a biased experimental design.

The correlation curves in \cref{fig:case02_corrs} show that all utility models remain positively correlated across the design domain, but the strength of these correlations varies substantially with the measurement time. The correlation between the high-fidelity utility and FNO-M is generally the largest, while the correlation between the high-fidelity utility and FNO-L drops more sharply near intermediate measurement times before increasing again toward later times. The FNO-M and FNO-L utilities are also correlated with one another, but this correlation varies across the domain. This variation indicates that a single domain-averaged covariance matrix is unlikely to provide uniformly optimal ACV hyperparameters.

The transformed covariance quantities in \cref{fig:case02_gammas,fig:case02_log_stdevs} show that the covariance structure varies smoothly but non-trivially over the design domain. The $\gamma$-parameters capture the design-dependent correlation structure, while the log standard deviations show that the utility-model variances decrease with later measurement times. These smooth trends make the problem well suited to covariance emulation: pilot samples collected at one measurement time can inform nearby times, but the emulator must still preserve local variation in both correlations and marginal variances.

These observations highlight the motivation for MF-EIG with covariance emulation. The low-fidelity models are not accurate enough to optimize directly, but they are sufficiently correlated with the high-fidelity utility to serve as useful control variates. The unbiasedness of MF-EIG with respect to the high-fidelity utility model is therefore essential, while the design dependence of the correlations and variances motivates adapting the ACV hyperparameters across the measurement-time domain.

As in \cref{ss:four-branch}, we use BO with \texttt{qLogNoisyExpectedImprovement} for the stochastic optimization part. The total computational budget is fixed at $B=37.1$ hours and the number of mean-optimization iterations is fixed at $N_{\mathrm{opt}}=20$. For \textbf{MFEIG-$\boldsymbol{\gamma}$-OPT}, the same total budget is used, but the method first performs $N_{\mathrm{pd}}=10$ active-learning pilot-design queries with a budget of $2.47$ hours each. The remaining budget is then distributed evenly over the $20$ mean-optimization iterations. Thus, \textbf{MFEIG-$\boldsymbol{\gamma}$-OPT} spends approximately one third of the total budget on pilot sampling, while \textbf{NMC-SF} spends all of its budget on mean-objective optimization.

\begin{figure}[htbp]
    \centering
    \includegraphics[width=0.7\linewidth]{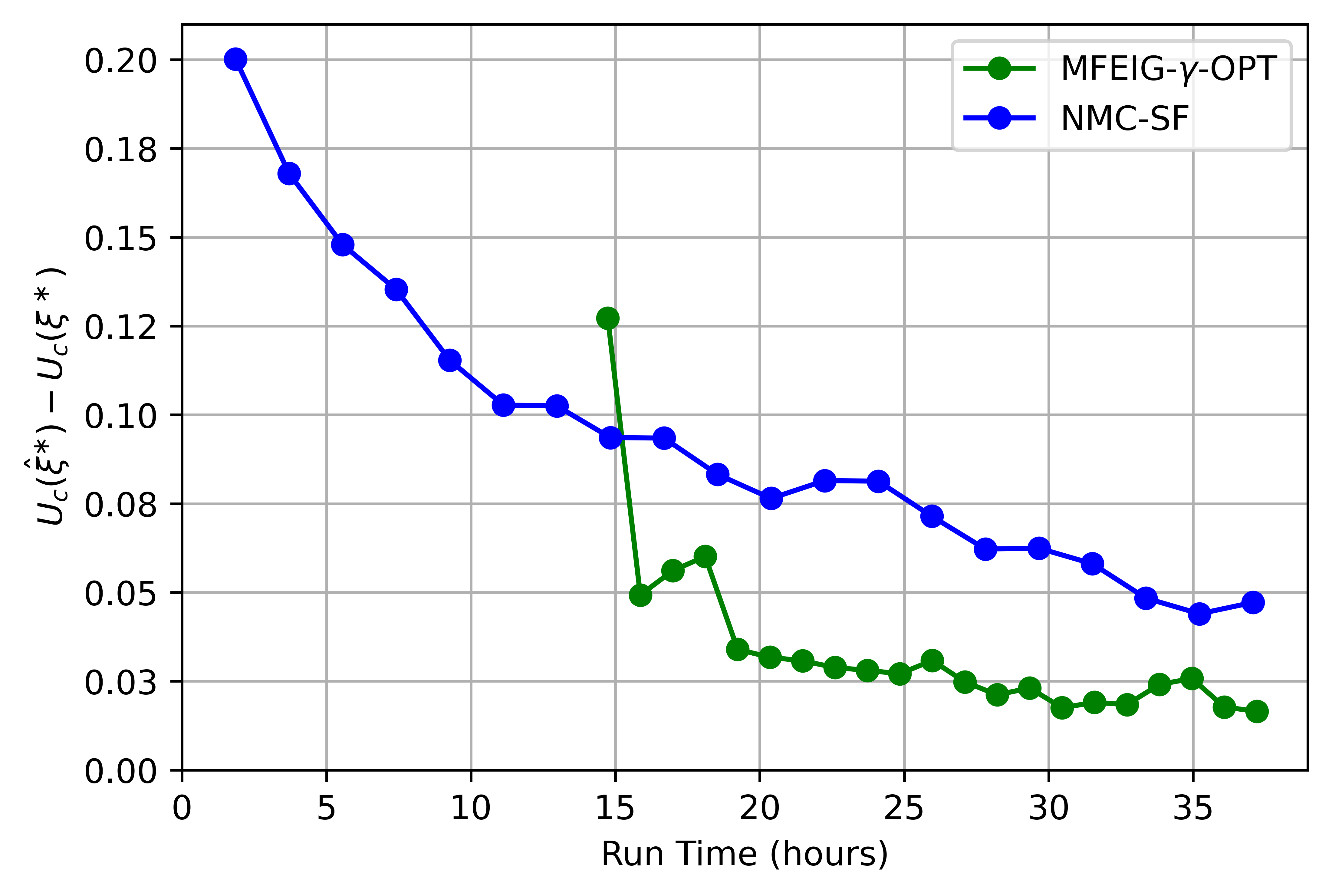}
    \caption{Optimization regrets for Case 2, the sensor measurement-time OED problem. The \textbf{MFEIG-$\boldsymbol{\gamma}$-OPT} curve begins after the active-learning pilot-sampling stage. Although this method spends a substantial portion of the budget on pilot sampling, it converges faster during mean optimization and reaches lower final regret than the \textbf{NMC-SF} baseline.}
    \label{fig:case2_regrets}
\end{figure}

Average regrets over runtime are shown in \cref{fig:case2_regrets}. The \textbf{MFEIG-$\boldsymbol{\gamma}$-OPT} curve begins after the pilot-sampling stage, so it initially has less remaining budget for objective optimization than \textbf{NMC-SF}. Despite this cost, the variance reduction from MF-EIG leads to faster convergence once the mean-optimization loop begins. By the end of the optimization procedure, \textbf{MFEIG-$\boldsymbol{\gamma}$-OPT} achieves roughly one third of the regret of \textbf{NMC-SF}. Together with Case 1, these results show that covariance-aware MF-EIG estimation can improve OED performance both in all-at-once grid-search settings and in iterative BO settings.

%% file: sections/5_conclusions.tex
\section{Conclusions and future work}
\label{s: conclusions}

Multi-fidelity estimators such as ACV can substantially reduce the cost of sampling-based objective estimation, but their performance depends on accurate model-output covariance information. In iterative stochastic optimization, this covariance information is generally design dependent. Estimating a new covariance matrix independently at every queried design can be expensive and noisy, while using a single domain-averaged covariance matrix can miss important local structure. This paper addressed this issue by formulating pilot sampling for multi-fidelity stochastic optimization as a covariance-emulation problem.

We introduced a probabilistic covariance emulator based on a positive-definite-preserving parameterization of the covariance matrix. The covariance is decomposed into standard deviations and correlations, with matrix-logarithm-based correlation parameters and log standard deviations modeled by independent GPs across the design domain. This construction allows pilot data to be shared across designs while still producing local covariance estimates for new optimization queries. We then developed active-learning acquisition functions that select pilot-design locations by targeting posterior uncertainty in the covariance emulator and emphasizing regions expected to be important for objective optimization.

The numerical demonstrations support the value of this approach. In the four-branch benchmark, \textbf{ACV-$\boldsymbol{\gamma}$-OPT} achieved the largest regret reductions at termination, outperforming single-fidelity MC and simpler pilot-sampling strategies. The results also show the tradeoff between global covariance sharing and local covariance adaptation: domain-averaged covariance estimation can perform well early by reducing covariance-estimation noise, while design-specific covariance information becomes more valuable as the optimizer concentrates near promising regions. In the Bayesian OED application, locally adapted MF-EIG estimators improved regret and estimator variance in the sensor-placement case, and the full adaptive pilot-sampling workflow improved convergence in the measurement-time case despite spending a substantial fraction of the total budget on pilot sampling.

Several limitations suggest directions for future work. First, this paper does not optimize the division of computational budget between pilot sampling and objective optimization; instead, it assumes a prescribed pilot-sampling budget and uses active learning to allocate it across the design domain. Jointly optimizing this budget split remains an important open problem. Second, the current emulator uses independent GPs for each transformed covariance parameter. Structured multi-output emulators or covariance-valued kernels may improve performance for larger model ensembles or higher-dimensional design domains. Third, the present implementation uses plug-in covariance estimates to determine ACV weights and sample allocations, with only a heuristic correction for covariance-emulator uncertainty. A more complete Bayesian treatment could propagate emulator uncertainty through the ACV design problem and the outer optimizer.

Finally, while this work focuses on ACV and MF-EIG estimators, the covariance-emulation perspective applies more broadly. Many multi-fidelity estimators, including MFMC, MLBLUE, and grouped ACV variants, require covariance or correlation information to set estimator hyperparameters. Extending active covariance learning to these estimators, as well as to constrained OUU and sequential or batch OED settings, would further broaden the applicability of the proposed framework.

%% file: sections/6_appendix.tex
\appendix

\section{Four-branch optimization problem results and details}\label{app:four_branch}

We provide here plots and results corresponding to the four-branch stochastic optimization problem from \cref{ss:four-branch}.

\subsection{Illustrative optimization run results}\label{app:four_branch_single}

The results below correspond to the single illustrative optimization run described in \cref{sss:results}.

\begin{figure}[htbp]
    \centering
    \includegraphics[width=0.9\linewidth]{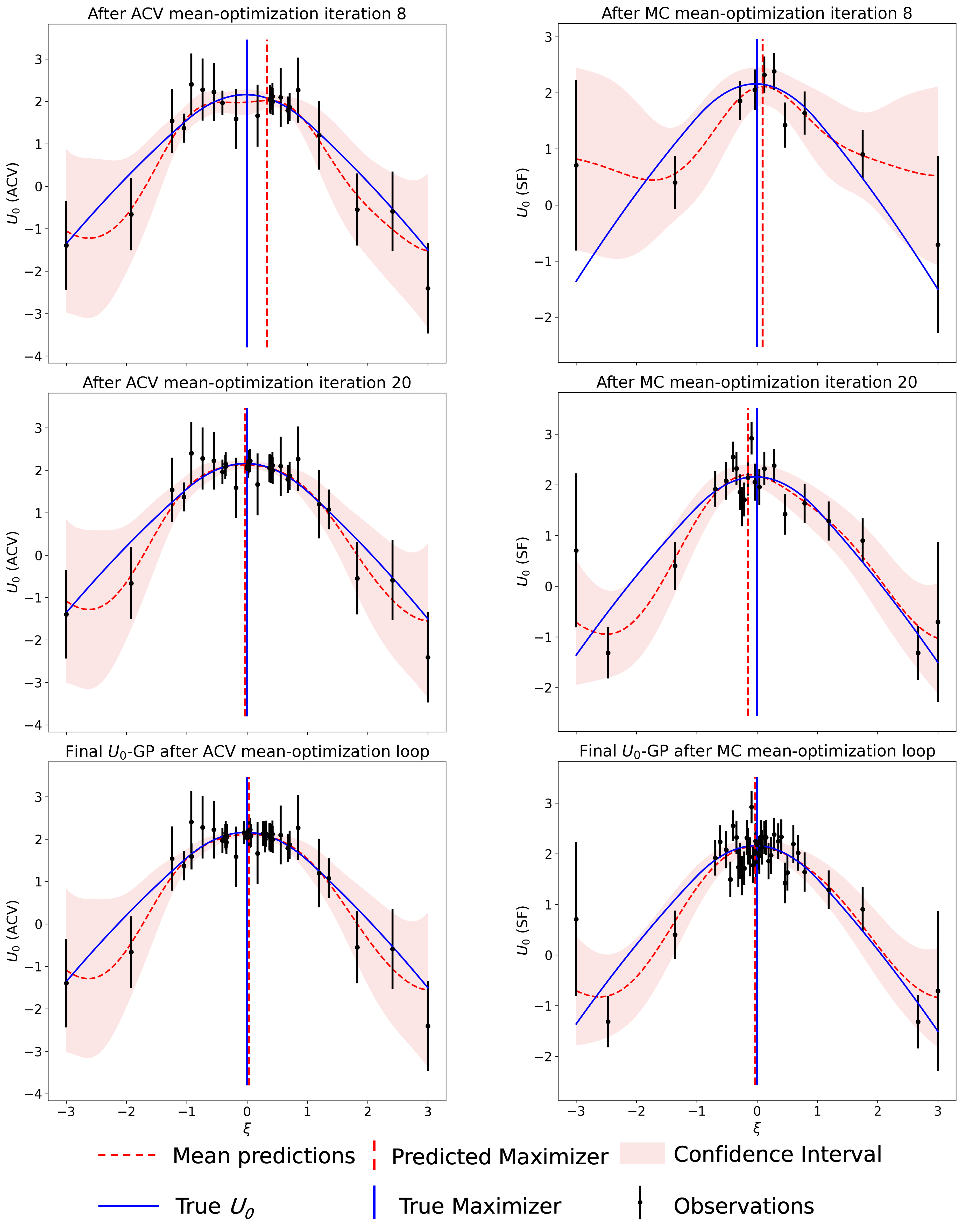}
    \caption{Comparison of the $U_0$-GP posteriors and predicted maximizers for one four-branch optimization run using \textbf{ACV-$\boldsymbol{\gamma}$-OPT} (left column) and \textbf{MC-SF} (right column). The rows show checkpoints after $8$, $20$, and $40$ mean-optimization iterations. The multi-fidelity \textbf{ACV-$\boldsymbol{\gamma}$-OPT} method produces substantially lower posterior uncertainty near the region of interest and converges more quickly to the true maximizer.}
    \label{fig:four_branch_demo}
\end{figure}

\clearpage

\subsection{Optimization results across all parameter configurations}\label{app:four_branch_results}

In \cref{ss:four-branch}, we summarized the four-branch benchmark results after averaging over multiple optimization hyperparameter settings. Here, we report the full regret results for each configuration. \Cref{tab:regret_results_np6} shows results for $N_{\mathrm{pd}}=6$ active-learning pilot-design locations, and \cref{tab:regret_results_np12} shows results for $N_{\mathrm{pd}}=12$. Each entry is the average optimization regret over $30$ independent trials. Within each configuration and budget checkpoint, the lowest average regret is shown in bold.

\begin{table}[htbp]
    \centering
    \caption{Average optimization regret at budget checkpoints for the four-branch benchmark with $N_{\mathrm{pd}}=6$. Each entry is averaged over $30$ independent trials; bold values indicate the lowest regret within each configuration and budget checkpoint.}
    \label{tab:regret_results_np6}
    \begin{tabular}{|c|l|c|c|c|}
        \hline
        \textbf{Setup} & \textbf{Method} & \textbf{20\% Budget} & \textbf{50\% Budget} & \textbf{100\% Budget} \\
        \hline\hline
        
        \multirow{4}{*}{\begin{tabular}{@{}l@{}}$B_{\mathrm{iter}}=32$\\$N_{\mathrm{pilot}}=16$\end{tabular}} 
        & MC-SF & $1.64 \times 10^{0}$ & $3.54 \times 10^{-1}$ & $4.98 \times 10^{-2}$ \\
        & ACV-$\boldsymbol{\gamma}$-OPT & $\mathbf{3.57 \times 10^{-2}}$ & $\mathbf{1.27 \times 10^{-2}}$ & $\mathbf{4.32 \times 10^{-3}}$ \\
        & ACV-FLAT & $5.00 \times 10^{-2}$ & $1.47 \times 10^{-2}$ & $1.15 \times 10^{-2}$ \\
        & ACV-INDEP & $9.84 \times 10^{-2}$ & $2.06 \times 10^{-2}$ & $7.59 \times 10^{-3}$ \\
        \hline

        \multirow{4}{*}{\begin{tabular}{@{}l@{}}$B_{\mathrm{iter}}=32$\\$N_{\mathrm{pilot}}=32$\end{tabular}} 
        & MC-SF & $1.69 \times 10^{0}$ & $2.95 \times 10^{-1}$ & $8.28 \times 10^{-2}$ \\
        & ACV-$\boldsymbol{\gamma}$-OPT & $5.35 \times 10^{-2}$ & $1.27 \times 10^{-2}$ & $\mathbf{5.79 \times 10^{-3}}$ \\
        & ACV-FLAT & $\mathbf{5.08 \times 10^{-2}}$ & $\mathbf{9.50 \times 10^{-3}}$ & $8.81 \times 10^{-3}$ \\
        & ACV-INDEP & $7.67 \times 10^{-2}$ & $2.17 \times 10^{-2}$ & $7.64 \times 10^{-3}$ \\
        \hline

        \multirow{4}{*}{\begin{tabular}{@{}l@{}}$B_{\mathrm{iter}}=64$\\$N_{\mathrm{pilot}}=16$\end{tabular}} 
        & MC-SF & $1.74 \times 10^{0}$ & $2.31 \times 10^{-1}$ & $7.36 \times 10^{-2}$ \\
        & ACV-$\boldsymbol{\gamma}$-OPT & $\mathbf{3.83 \times 10^{-2}}$ & $\mathbf{1.37 \times 10^{-2}}$ & $\mathbf{1.67 \times 10^{-3}}$ \\
        & ACV-FLAT & $6.84 \times 10^{-2}$ & $2.08 \times 10^{-2}$ & $9.20 \times 10^{-3}$ \\
        & ACV-INDEP & $1.48 \times 10^{-1}$ & $2.34 \times 10^{-2}$ & $7.43 \times 10^{-3}$ \\
        \hline

        \multirow{4}{*}{\begin{tabular}{@{}l@{}}$B_{\mathrm{iter}}=64$\\$N_{\mathrm{pilot}}=32$\end{tabular}} 
        & MC-SF & $1.65 \times 10^{0}$ & $2.38 \times 10^{-1}$ & $5.25 \times 10^{-2}$ \\
        & ACV-$\boldsymbol{\gamma}$-OPT & $3.73 \times 10^{-2}$ & $\mathbf{5.75 \times 10^{-3}}$ & $\mathbf{3.13 \times 10^{-3}}$ \\
        & ACV-FLAT & $\mathbf{1.54 \times 10^{-2}}$ & $1.13 \times 10^{-2}$ & $8.43 \times 10^{-3}$ \\
        & ACV-INDEP & $9.82 \times 10^{-2}$ & $1.60 \times 10^{-2}$ & $5.98 \times 10^{-3}$ \\
        \hline

        \multirow{4}{*}{\begin{tabular}{@{}l@{}}$B_{\mathrm{iter}}=128$\\$N_{\mathrm{pilot}}=16$\end{tabular}} 
        & MC-SF & $1.51 \times 10^{0}$ & $3.47 \times 10^{-1}$ & $3.11 \times 10^{-2}$ \\
        & ACV-$\boldsymbol{\gamma}$-OPT & $\mathbf{2.90 \times 10^{-2}}$ & $\mathbf{4.71 \times 10^{-3}}$ & $\mathbf{2.22 \times 10^{-3}}$ \\
        & ACV-FLAT & $9.14 \times 10^{-2}$ & $2.30 \times 10^{-2}$ & $6.81 \times 10^{-3}$ \\
        & ACV-INDEP & $1.30 \times 10^{-1}$ & $1.79 \times 10^{-2}$ & $9.46 \times 10^{-3}$ \\
        \hline

        \multirow{4}{*}{\begin{tabular}{@{}l@{}}$B_{\mathrm{iter}}=128$\\$N_{\mathrm{pilot}}=32$\end{tabular}} 
        & MC-SF & $1.61 \times 10^{0}$ & $3.50 \times 10^{-1}$ & $3.75 \times 10^{-2}$ \\
        & ACV-$\boldsymbol{\gamma}$-OPT & $2.23 \times 10^{-2}$ & $\mathbf{3.53 \times 10^{-3}}$ & $\mathbf{1.15 \times 10^{-3}}$ \\
        & ACV-FLAT & $\mathbf{8.61 \times 10^{-3}}$ & $8.07 \times 10^{-3}$ & $4.67 \times 10^{-3}$ \\
        & ACV-INDEP & $6.83 \times 10^{-2}$ & $8.93 \times 10^{-3}$ & $4.89 \times 10^{-3}$ \\
        \hline
    \end{tabular}
\end{table}

\begin{table}[htbp]
    \centering
    \small
    \setlength{\tabcolsep}{3.5pt}
    \caption{Average optimization regret at budget checkpoints for the four-branch benchmark with $N_{\mathrm{pd}}=12$. Each entry is averaged over $30$ independent trials; bold values indicate the lowest regret within each configuration and budget checkpoint.}
    \label{tab:regret_results_np12}
    \resizebox{\linewidth}{!}{\begin{tabular}{|c|l|c|c|c|}
        \hline
        \textbf{Configuration} & \textbf{Method} & \textbf{20\% Budget} & \textbf{50\% Budget} & \textbf{100\% Budget} \\
        \hline\hline
        
        \multirow{4}{*}{\begin{tabular}{@{}l@{}}$B_{\mathrm{iter}}=32$\\$N_{\mathrm{pilot}}=16$\end{tabular}} 
        & MC-SF & $1.69 \times 10^{0}$ & $2.95 \times 10^{-1}$ & $8.28 \times 10^{-2}$ \\
        & ACV-$\boldsymbol{\gamma}$-OPT & $\mathbf{6.52 \times 10^{-2}}$ & $\mathbf{8.17 \times 10^{-3}}$ & $\mathbf{3.58 \times 10^{-3}}$ \\
        & ACV-FLAT & $9.87 \times 10^{-2}$ & $1.96 \times 10^{-2}$ & $1.42 \times 10^{-2}$ \\
        & ACV-INDEP & $1.03 \times 10^{-1}$ & $2.45 \times 10^{-2}$ & $7.95 \times 10^{-3}$ \\
        \hline

        \multirow{4}{*}{\begin{tabular}{@{}l@{}}$B_{\mathrm{iter}}=32$\\$N_{\mathrm{pilot}}=32$\end{tabular}} 
        & MC-SF & $1.57 \times 10^{0}$ & $3.23 \times 10^{-1}$ & $9.24 \times 10^{-2}$ \\
        & ACV-$\boldsymbol{\gamma}$-OPT & $\mathbf{6.84 \times 10^{-2}}$ & $\mathbf{8.22 \times 10^{-3}}$ & $\mathbf{2.53 \times 10^{-3}}$ \\
        & ACV-FLAT & $7.03 \times 10^{-2}$ & $1.12 \times 10^{-2}$ & $5.16 \times 10^{-3}$ \\
        & ACV-INDEP & $1.30 \times 10^{-1}$ & $1.57 \times 10^{-2}$ & $6.13 \times 10^{-3}$ \\
        \hline

        \multirow{4}{*}{\begin{tabular}{@{}l@{}}$B_{\mathrm{iter}}=64$\\$N_{\mathrm{pilot}}=16$\end{tabular}} 
        & MC-SF & $1.65 \times 10^{0}$ & $2.38 \times 10^{-1}$ & $5.25 \times 10^{-2}$ \\
        & ACV-$\boldsymbol{\gamma}$-OPT & $\mathbf{1.78 \times 10^{-2}}$ & $\mathbf{3.42 \times 10^{-3}}$ & $\mathbf{2.07 \times 10^{-3}}$ \\
        & ACV-FLAT & $3.96 \times 10^{-2}$ & $1.72 \times 10^{-2}$ & $5.24 \times 10^{-3}$ \\
        & ACV-INDEP & $1.20 \times 10^{-1}$ & $1.90 \times 10^{-2}$ & $7.99 \times 10^{-3}$ \\
        \hline

        \multirow{4}{*}{\begin{tabular}{@{}l@{}}$B_{\mathrm{iter}}=64$\\$N_{\mathrm{pilot}}=32$\end{tabular}} 
        & MC-SF & $1.58 \times 10^{0}$ & $2.73 \times 10^{-1}$ & $3.42 \times 10^{-2}$ \\
        & ACV-$\boldsymbol{\gamma}$-OPT & $\mathbf{2.36 \times 10^{-2}}$ & $\mathbf{5.25 \times 10^{-3}}$ & $\mathbf{1.50 \times 10^{-3}}$ \\
        & ACV-FLAT & $2.90 \times 10^{-2}$ & $7.24 \times 10^{-3}$ & $3.95 \times 10^{-3}$ \\
        & ACV-INDEP & $1.51 \times 10^{-1}$ & $1.68 \times 10^{-2}$ & $6.77 \times 10^{-3}$ \\
        \hline

        \multirow{4}{*}{\begin{tabular}{@{}l@{}}$B_{\mathrm{iter}}=128$\\$N_{\mathrm{pilot}}=16$\end{tabular}} 
        & MC-SF & $1.61 \times 10^{0}$ & $3.50 \times 10^{-1}$ & $3.75 \times 10^{-2}$ \\
        & ACV-$\boldsymbol{\gamma}$-OPT & $\mathbf{1.45 \times 10^{-2}}$ & $\mathbf{1.87 \times 10^{-3}}$ & $\mathbf{1.44 \times 10^{-3}}$ \\
        & ACV-FLAT & $2.60 \times 10^{-2}$ & $6.47 \times 10^{-3}$ & $2.28 \times 10^{-3}$ \\
        & ACV-INDEP & $9.68 \times 10^{-2}$ & $2.20 \times 10^{-2}$ & $7.07 \times 10^{-3}$ \\
        \hline

        \multirow{4}{*}{\begin{tabular}{@{}l@{}}$B_{\mathrm{iter}}=128$\\$N_{\mathrm{pilot}}=32$\end{tabular}} 
        & MC-SF & $1.64 \times 10^{0}$ & $3.34 \times 10^{-1}$ & $2.96 \times 10^{-2}$ \\
        & ACV-$\boldsymbol{\gamma}$-OPT & $1.24 \times 10^{-2}$ & $5.44 \times 10^{-3}$ & $\mathbf{1.45 \times 10^{-3}}$ \\
        & ACV-FLAT & $\mathbf{1.05 \times 10^{-2}}$ & $\mathbf{3.36 \times 10^{-3}}$ & $2.06 \times 10^{-3}$ \\
        & ACV-INDEP & $3.74 \times 10^{-2}$ & $2.10 \times 10^{-2}$ & $4.92 \times 10^{-3}$ \\
        \hline
    \end{tabular}}
\end{table}

\clearpage

\section{Advection-diffusion OED reference quantities}\label{app:ad-vec-ref}

We provide below the reference solutions and correlation quantities for both cases of the advection-diffusion OED problem from \cref{ss:OED}.

\subsection{Case 1: OED sensor placement optimization}\label{app:advec-ref-case-1}

Below is the reference EIG objective surface across different possible sensor-placement locations, with the star indicating the optimal sensor placement.

\begin{figure}[htbp]
    \centering
    \includegraphics[width=0.55\linewidth]{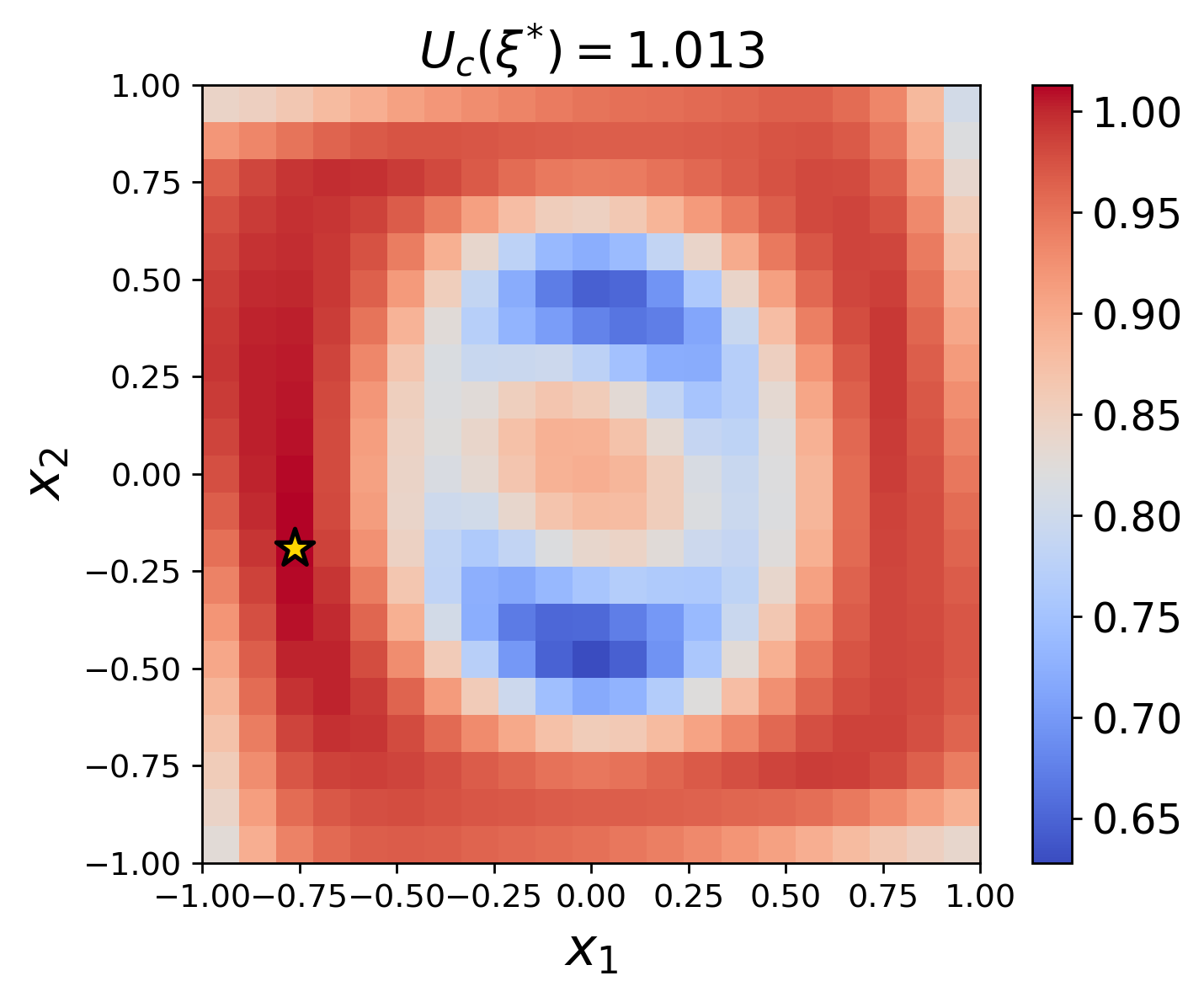}
    \caption{Reference EIG surface for the OED sensor-placement problem in Case 1. Larger EIG values correspond to more informative sensor locations, and the reference optimum is marked at $\design^\ast=[-0.8,-0.2]^\top$.}
    \label{fig:reference_util}
\end{figure}

\clearpage

\subsection{Case 2: OED sensor measurement-time optimization}\label{app:advec-ref-case-2}

Below is the reference cost-aware utilities and covariance quantities (correlations, correlation parameters, and log standard deviations) across the design domain of possible sensor times.

\begin{figure}[htbp]
    \centering
    \begin{subfigure}[b]{0.49\textwidth}
        \centering
        \includegraphics[width=0.9\linewidth]{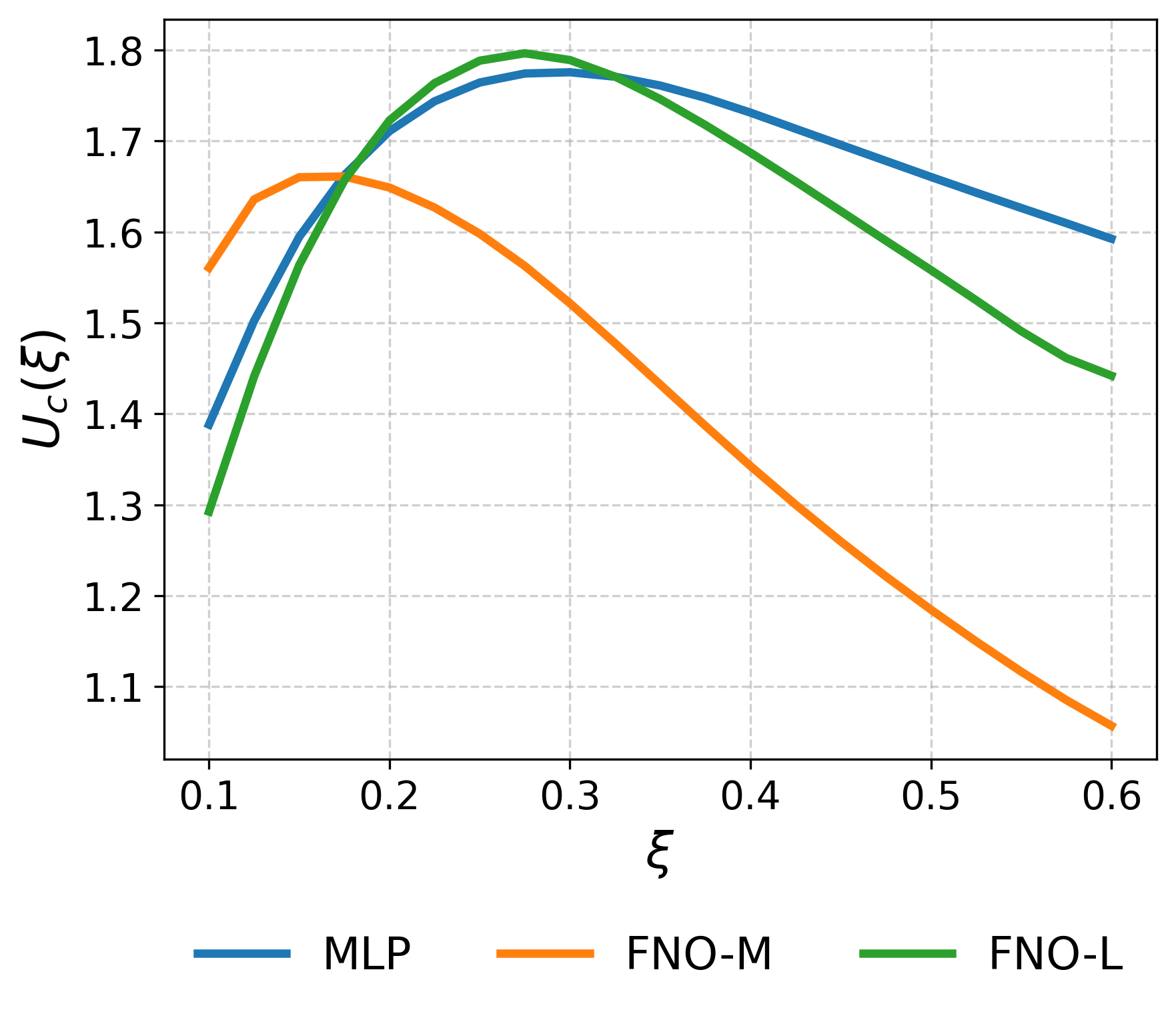}
        \caption{Cost-aware utility means}
        \label{fig:case02_pilot_utils}
    \end{subfigure}
    \hfill 
    \begin{subfigure}[b]{0.49\textwidth}
        \centering
        \includegraphics[width=0.9\linewidth]{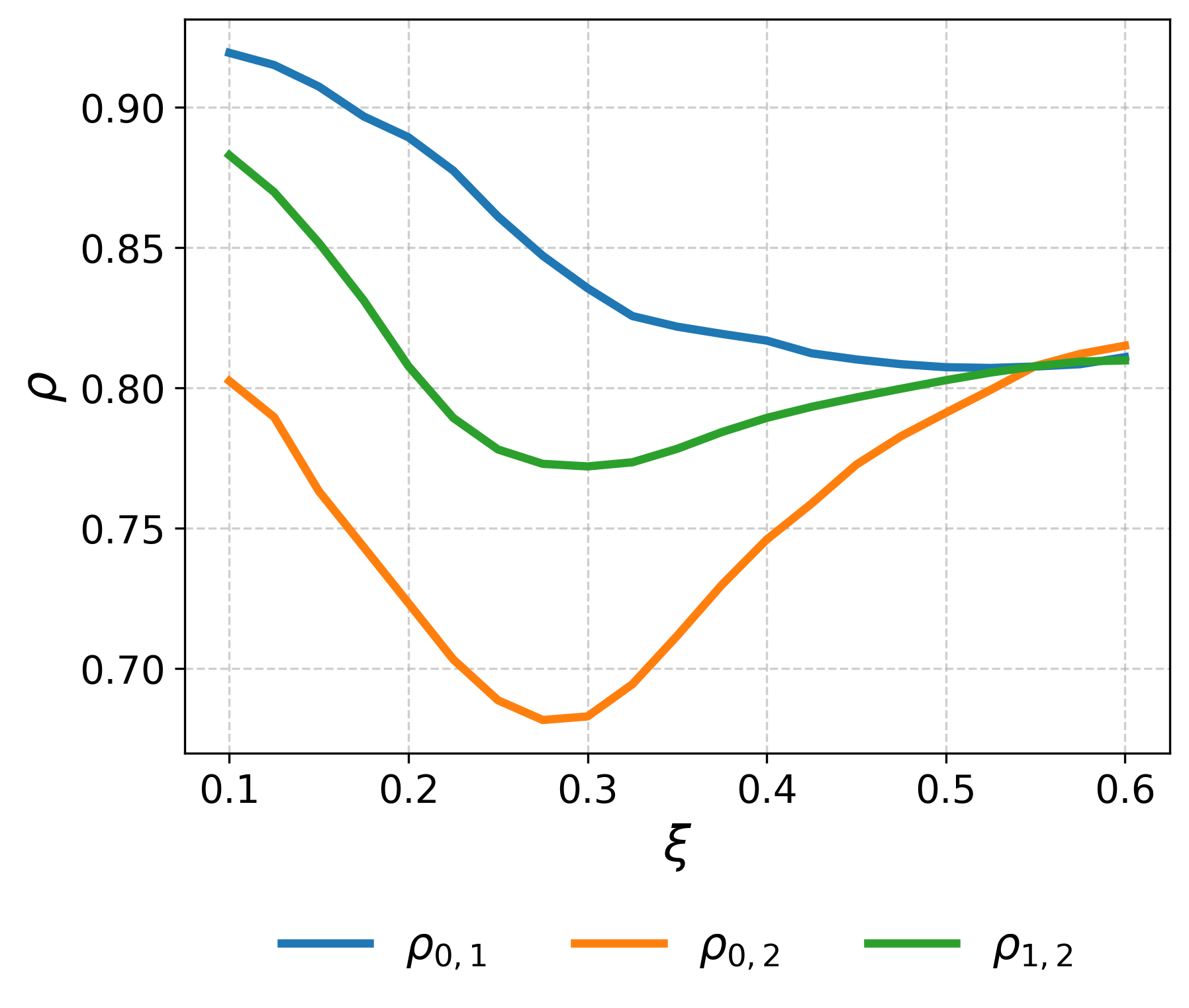}
        \caption{Utility-model correlations}
        \label{fig:case02_corrs}
    \end{subfigure}
    \vspace{0.2em}
    \begin{subfigure}[b]{0.49\textwidth}
        \centering
        \includegraphics[width=0.9\linewidth]{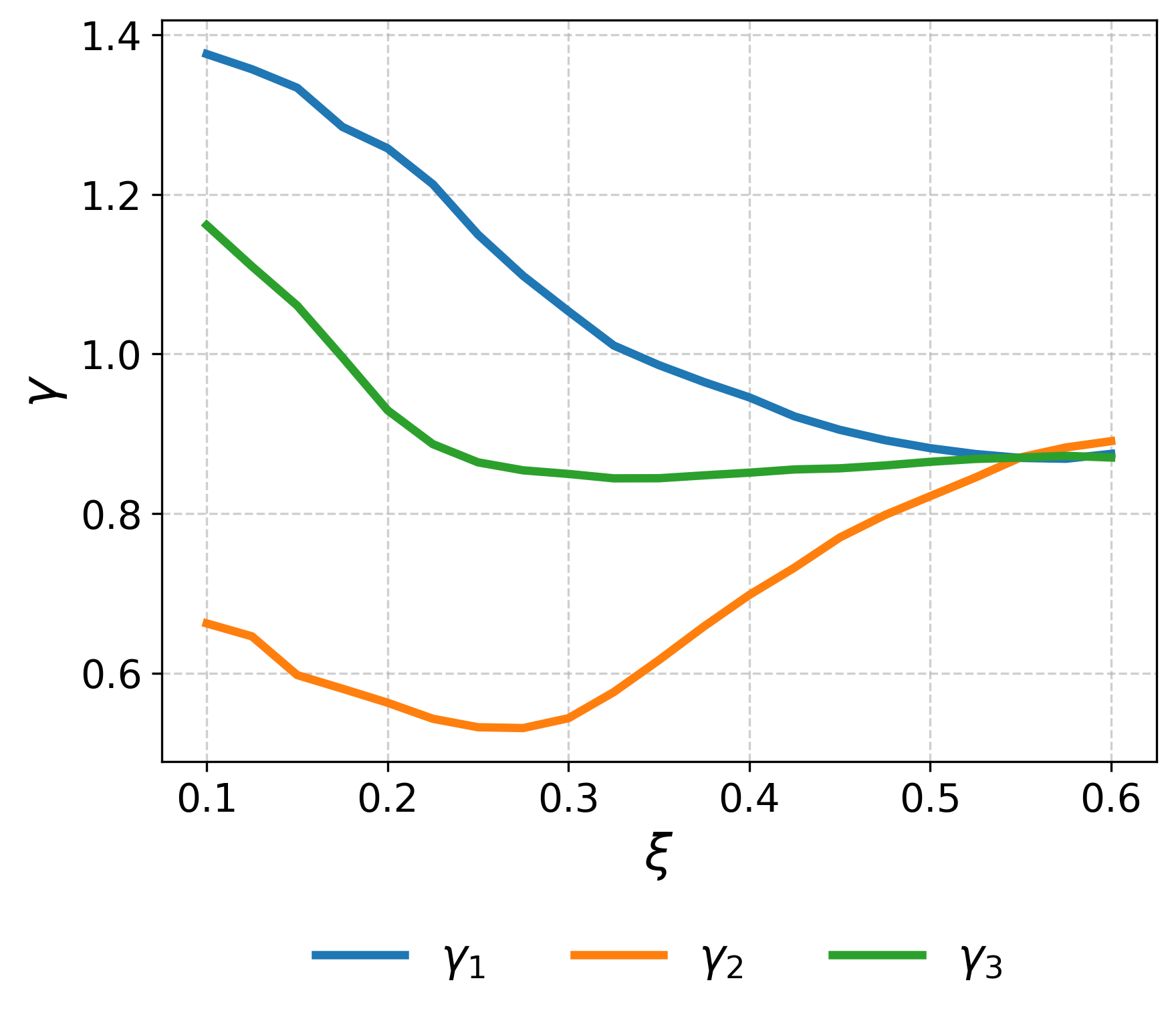}
        \caption{Correlation parameters $\gamma(\design)$}
        \label{fig:case02_gammas}
    \end{subfigure}
    \hfill
    \begin{subfigure}[b]{0.49\textwidth}
        \centering
        \includegraphics[width=0.9\linewidth]{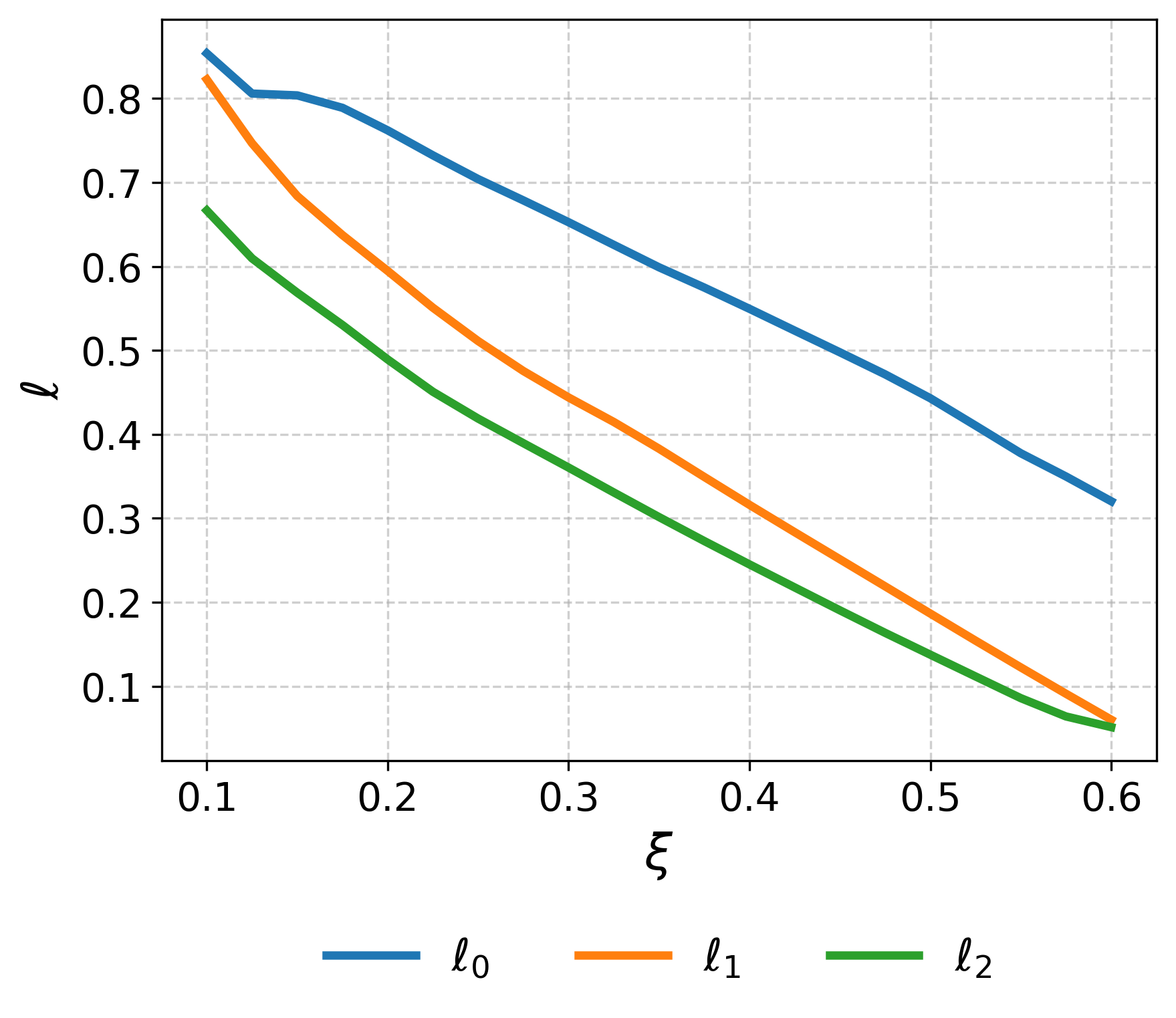}
        \caption{Log standard deviations $\ell(\design)$}
        \label{fig:case02_log_stdevs}
    \end{subfigure}
    \caption{Reference quantities for Case 2, the sensor measurement-time optimization problem. \textbf{(a)} Cost-aware utility means for each utility model. \textbf{(b)} Utility-model correlations across the design domain. \textbf{(c)} Correlation parameters $\gamma(\design)$ obtained from the bijective covariance parameterization. \textbf{(d)} Log standard deviations $\ell(\design)$ of the utility models. }
    \label{fig:case02_pilot_reference}
\end{figure}

\clearpage

\section{Advection-diffusion model details}\label{app:ad-vec-models}

Here we provide further results and plots related to the advection-diffusion multi-fidelity model ensemble and the proxy model used in lieu of the full model in our demonstrations for \cref{sss:oed-results}.

\subsection{Multi-fidelity model ensemble realizations}\label{app:ad-vec-model-realizations}

We compare two different runs of the multi-fidelity model ensemble based on two realizations from the uncertain parameters $\Theta$.

\begin{figure}[htbp]
    \centering
    \includegraphics[width=\linewidth]{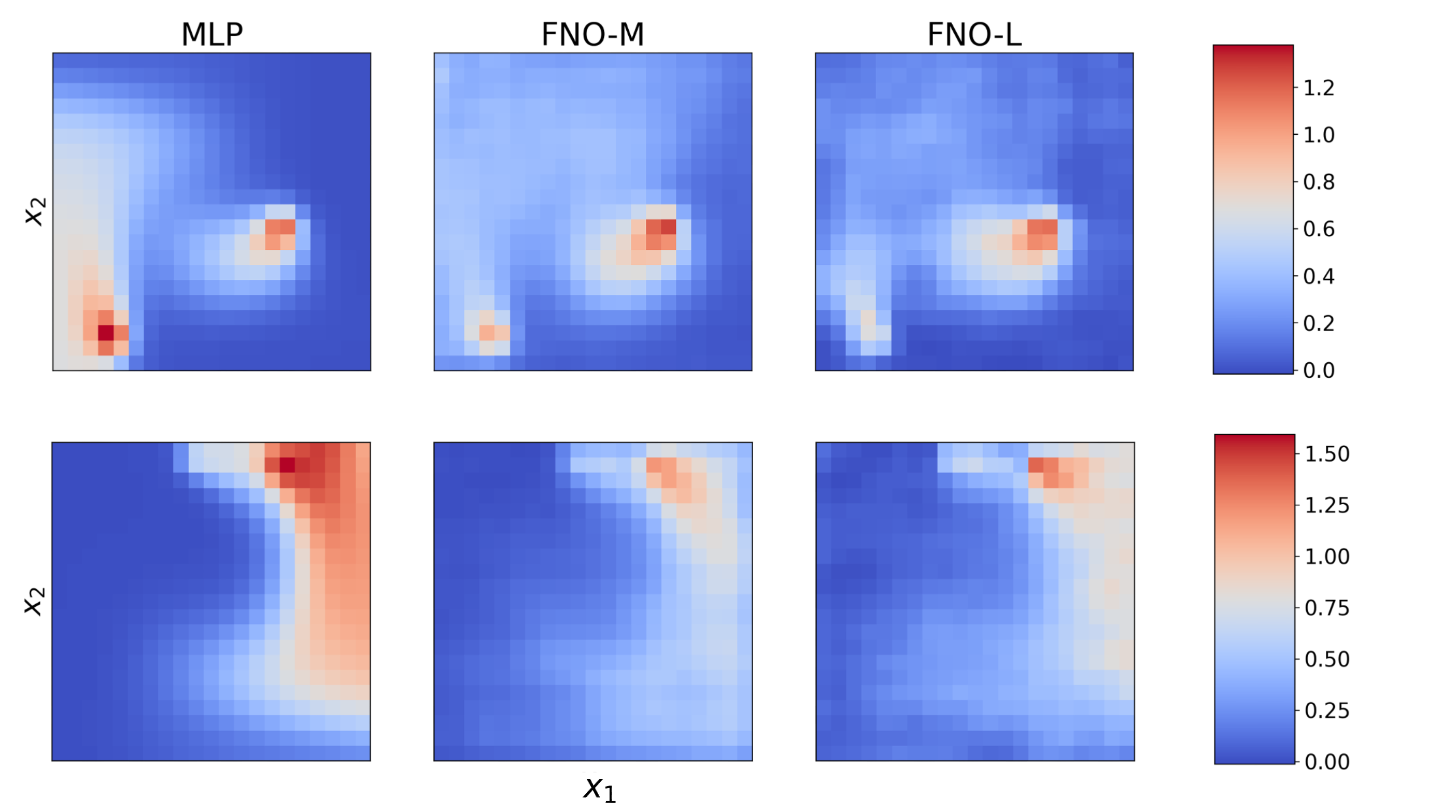}
    \caption{Examples of contaminant concentration fields at $t_k=0.325$. Each row corresponds to a different realization of $\Theta$. The high-fidelity proxy model $G_0$ (left) is compared with the medium-fidelity FNO $G_1$ (middle) and low-fidelity FNO $G_2$ (right).}
    \label{fig:mlp_fno_fnolf_comparison}
\end{figure}

\clearpage

\subsection{Proxy model details}
\label{app:proxy}

The training data are generated by solving \eqref{eq:conv_diff_2d_temporal} for $400$ independent samples of $\Theta$ drawn from the prior distribution. The MLP maps directly from spatio-temporal coordinates and uncertain parameters to the concentration value. Its input is
\begin{align}
    \left(
        x_1,
        x_2,
        t,
        \Theta_{1,x},
        \Theta_{1,y},
        \Theta_{2,x},
        \Theta_{2,y},
        \Theta_{\beta}
    \right),
\end{align}
and its scalar output is the predicted concentration $G(x,t;\Theta)$. Unlike the finite-difference solver, which must march the full state forward in time, the MLP provides direct pointwise predictions of the concentration field. The network has five hidden layers with $48$, $64$, $128$, $48$, and $10$ nodes, respectively, with nonlinear activation functions between hidden layers.

\begin{figure}[htbp]
    \centering
    \includegraphics[width=0.8\linewidth]{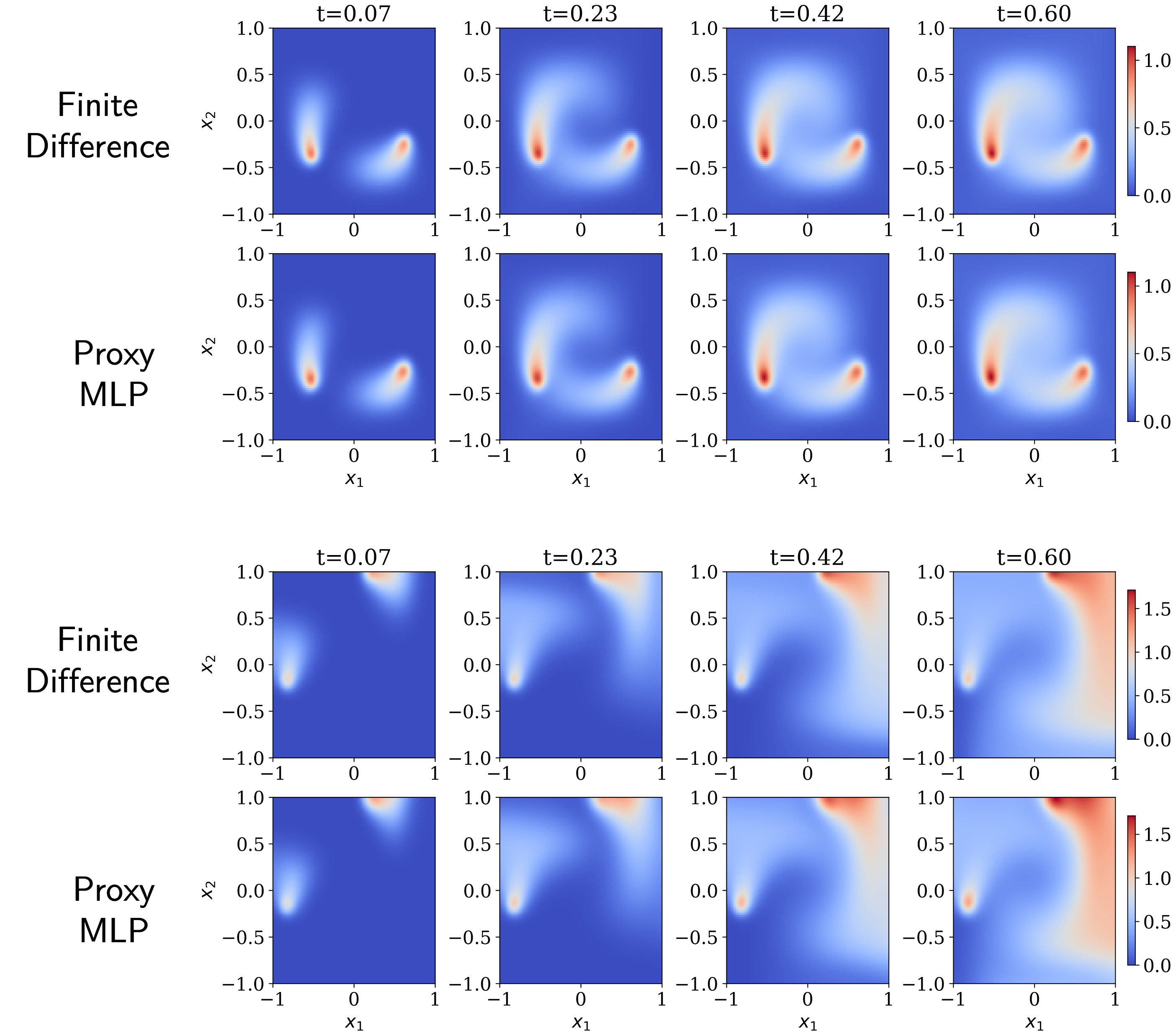}
    \caption{Comparison between the finite-difference solution of \eqref{eq:conv_diff_2d_temporal} and the MLP proxy for representative source-parameter realizations and times. The MLP captures the concentration plume induced by the source term $S(x;\Theta)$ and transported by the spiral-like velocity field $v(x;\Theta)$.}
    \label{fig:hifi_mlp_comparison}
\end{figure}

\Cref{fig:hifi_mlp_comparison} compares concentration fields produced by the finite-difference solver and the MLP proxy for representative parameter realizations and times. The MLP accurately captures the plume shape and its counterclockwise transport under the spiral-like velocity field, providing a sufficiently accurate proxy for constructing reference EIG surfaces and benchmarking the proposed stochastic optimization workflow.

\clearpage